\documentclass{emulateapj}

\shorttitle{NCs and BHs}
\shortauthors{Neumayer, N., Walcher, C.J.}

\def\msun{$\mbox{M}_\odot$} 
\def\mbh{$\mbox{M}_{\mathrm{BH}}$ }
\def\mnc{$\mbox{M}_{\mathrm{NC}}$ }
\def\mbulge{$\mbox{M}_{\mathrm{bulge}}$ }
\def\nsersic{$\mbox{n}_{\mathrm{Sersic}}$ }
\def\kms{$\mbox{km s}^{-1}$}

\begin{document}

%\title{Black holes in nuclear star clusters:\\
%mass estimates and scaling relations}

\title{Are nuclear star clusters the precursors of massive black holes?}

\author{Nadine Neumayer}
\affil{European Southern Observatory, Karl-Schwarzschild Str. 2, 85748 Garching bei M\"unchen, Germany\\
Excellence Cluster Universe, Boltzmann Str. 2, 85748 Garching bei M\"unchen, Germany}
\email{nneumaye@eso.org}
\author{C. Jakob Walcher}
\affil{Leibniz-Institut f\"ur Astrophysik Potsdam (AIP), An der Sternwarte 16, 14482 Potsdam, Germany}
\email{jwalcher@aip.de}

% ABSTRACT
\begin{abstract}

We present new upper limits for black hole masses in extremely late type spiral galaxies. 
We confirm that this class of galaxies has black holes with masses less than $10^6 M_{\odot}$, 
if any. We also derive new upper limits for nuclear star cluster masses in massive galaxies 
with previously determined black hole masses. We use the newly derived upper limits and a 
literature compilation to study the low mass end 
of the global-to-nucleus relations. We find the following: 1) The \mbh - $\sigma$ relation cannot flatten at low 
masses, but may steepen. 2) The \mbh - \mbulge relation may well flatten in contrast. 3) The 
\mbh - Sersic n relation is able to account for the large scatter 
in black hole masses in low-mass disk galaxies. Outliers in the \mbh - Sersic n relation seem to 
be dwarf elliptical galaxies, which may imply that while the morphological transformation mechanism 
for massive galaxies is associated with black hole growth, this is not the case in dwarf galaxies. 
When plotting \mbh vs. \mnc we find three different regimes: 
a) nuclear cluster dominated nuclei, b) a transition region and c) black hole dominated nuclei. 
This is consistent with the picture, in which black holes form inside nuclear clusters with a very 
low-mass fraction. They subsequently grow much faster than the nuclear cluster, destroying it 
when the ratio \mbh / \mnc grows above 100. Nuclear star clusters 
may thus be the precursors of massive black holes in galaxy nuclei. \\

\end{abstract}

%INTRODUCTION
\section{Introduction}
Supermassive black holes (BHs) are thought to be ubiquitous in the nuclei of massive galaxies.
The discovery of a number of tight correlations between the global properties of galaxies and the
properties of their nuclei \citep[e.g.][]{ferrarese00, gebhardt00a, haring04} has led astronomers 
to realize that the evolution of galaxies may be closely linked to their nuclear properties. 
However, the nuclei of galaxies do not only host massive BHs, 
but also massive star clusters, commonly called nuclear star clusters (NC)\footnote{Note that we here make 
the distinction between nucleus, i.e. the location at the very center, and nuclear star cluster. Often the NC has been 
called nucleus or stellar nucleus in the past, but this seems ambiguous to us.}.
The overall nucleation frequency is around 75\% over all Hubble types 
\citep[][hereafter B02]{carollo98, cote06, boker02}, but NCs seem to be absent in the most 
massive galaxies \citep{graham09b, cote06}. 
NCs typically have stellar velocity dispersions of $15 - 35 $ \kms,
effective radii of a few parsecs, and dynamical masses of 
$\sim 10^6 - 10^7 M_{\odot}$ \citep[B02;][]{boker04, walcher05}.
Moreover, they show stellar populations of multiple ages \citep{rossa06,seth06,walcher06}, 
pointing towards them having a complex formation history. This might be related to their 
special location in the galaxy, as on average, NCs appear to sit at the photometric centre 
of their host galaxy \citep{binggeli00, boker02}. We recently showed that for bulgeless galaxies their location also coincides with the kinematic centre, i.e. the bottom of the potential well 
\citep{neumayer11}. 

Intriguingly, NCs in late-type spirals and dwarf ellipticals follow relationships with
their host galaxies that mirror the \mbh$ - \sigma$ and \mbh$ - $ \mbulge\ relationships of high-mass
galaxies \citep{rossa06, ferrarese06a, wehner06}, suggesting the
possibility that the fueling and growth of NCs and BHs are determined by similar processes, and that 
BHs and NCs should be grouped together into ``central massive objects" \citep[CMOs, ][]{ferrarese06a}. 
The NC would then be nothing else than the failed BH \citep{elmegreen08}. In this picture, BHs would 
form in high density clumps typically located in high mass galaxies, while NCs form from lower density 
clumps in lower density disks. Recent simulation studies \citep[e.g.][]{regan09,mayer10} have been able 
to reproduce the formation of BHs through direct collapse models. If the collapse is quick  - compared to 
the cooling time of the gas -  a BH will form. If however, the gas has sufficient time to cool and form stars, it will 
form a NC \citep[see also the recent review by ][]{bromm11}. Competing formation scenarios for NCs are, 
however, equally successful. For example recent work by \cite{hartmann11} has shown that the observed 
properties of NCs are well reproduced by combining mergers of star clusters with the accretion of gas at a 
much later time in the history of a galaxy.

A further reason for interest in NCs and their BHs is that a number of authors \citep{ebisuzaki01, portegies04, 
gurkan04, freitag06a, freitag06b, gaburov08} have found that dense clusters of young, massive stars can 
experience runaway coalescence of their most massive stars, leading to an intermediate mass 
black hole \citep[IMBH, but see also][]{glebbeek09}. It would then be tempting to identify NCs with the 
long-sought seeds for BH formation. An observational result supporting this view, is that NCs and BHs can 
coincide \citep{filippenko03, seth08}, this is especially well-studied in our own Galaxy \citep{schoedel07, genzel10}. 
On the other hand, parameter studies of the runaway collapse scenario \citep[e.g.][]{freitag06a} show that 
NCs are actually not in a region of parameter space that would be favourable to the collapse. 

Of the many global-to-nucleus relations, the three most frequently referred to ones seem to be the \mbh$-$ $\sigma$ relation 
\citep[][]{ferrarese00, gebhardt00a}, the \mbh$-$ \mbulge relation \citep[][]{haring04} 
and the \mbh$-$ \nsersic\ relation \citep{graham07}. As all these relations have been initially set up for the range 
of massive galaxies (i.e. \mbh$> 10^8$ \msun), the low-mass range of BHs is not very well populated and holds 
most potential to find out which one of the three
is more fundamental. A particularly interesting case are BHs and NCs in bulgeless galaxies. 
Indeed, while according to the \mbh$ - \sigma$ relation one would expect late-type, bulgeless spirals to host
BHs of mass $\le 10^6$\msun\ , the \mbh$ - $ \mbulge relation is no longer ``defined" for bulgeless galaxies. As the lack of a bulge would imply the absence of a black hole. On the other hand, exploring the low mass end of the scaling 
relations, \cite{greene10} have derived reliable BH 
masses in spiral galaxies (with bulges) from maser measurements and find that these fall below the \mbh$-$ $\sigma$ 
relation of elliptical galaxies, but seem consistent with the \mbh$ - $ \mbulge relation. 

In fact both NCs and BHs have been found in bulgeless galaxies. For NCs see B02, for BHs see e.g.~the cases of 
NGC4395 \citep{filippenko89, filippenko03}, NGC1042 \citep{shields08}, NGC3621 \citep{barth09, gliozzi09} and probably 
many more \citep[see e.g. ][]{satyapal08, greene07, barth08,mcalpine11}. On the other hand, very tight upper limits 
for the BH mass exist for some galaxies such as M33 \citep{gebhardt01, merritt01}, but direct 
observational constraints are scarce because such small BHs are extremely difficult targets for
dynamical searches and therefore very few objects have useful measurements.  While it would thus seem tempting to declare that NCs are the 
central spheroids in bulgeless galaxies, this could lead to a paradox. Indeed, NCs have largely been identified 
with CMOs in massive galaxies, on the ground that they follow similar scaling relations as BHs. Identifying 
the same objects with the spheroid in low-mass galaxies would imply a transition in physical properties of the 
NC. Many observational hints seem to point against this possibility  \citep{walcher05}, the most important being 
that NCs have constant radius over Hubble type. A backdoor might be that \cite{erwin10} find that BH mass correlates 
with bulge mass \citep[and no correlation with disk mass exists,][]{kormendy11a}, while NC mass correlates 
better with total galaxy mass. 

To conclude this introduction, measurements of the demographics of the lowest-mass BHs are
an important goal. Their mass distribution encodes a fossil record of the mass scale and
formation efficiency of the initial BH ÒseedsÓ at high redshift \citep[e.g.][]{volonteri08} and they 
hold the power to help us distinguish between different scenarios explaining the observed 
global-to-nucleus relations \citep[][]{dimatteo05, hopkins06, peng07, jahnke11}. 
In order to increase the statistics in this particularly interesting 
low-mass regime, we have calculated \mbh upper limits for a sample of 9 NCs, for which integrated velocity 
dispersions had been published previously \citep[][hereafter W05]{walcher05}. We have also placed upper limits 
on \mnc for a sample of 11 galaxies with measured black hole masses. We have then used these 
upper limits in conjunction with a literature compilation to gauge which of the different proposed 
global-to-nucleus relations seem to hold best at the low mass end.

\section{New upper limits for BHs in NCs}

\subsection{Data} 

Our sample consists of 9 NCs culled from the HST/WFPC2 snapshot survey of B02. 
Imaging in the F804W filter is available from B02 and we refer to this paper for all details. 
All 9 NCs are resolved, even if some only barely. We here use the images as available through MAST 
to yield the surface brightness profile through a multi-Gaussian expansion (see below). 

VLT/UVES spectra with high S/N and high spectral resolution have been obtained by W05. 
We use their velocity dispersion measurement. 
The properties of our sample are summarized in Table \ref{t:values}. 

The sample selection for spectroscopic follow-up technically implied a slight bias 
to the more luminous among the NCs. Nevertheless, we expect this sample to be a fair 
representation of NCs in pure disk galaxies in general, as it covers the upper 2/3 of the 
luminosity range of NCs.

\subsection{Analysis} 

We constructed a dynamical model to estimate the mass and M/L of the NCs
and to put meaningful upper limits on the possible central black holes inside them. The first
step in this process is developing a model for the light distribution. To parametrise the 
surface brightness profiles of the NCs and to deproject the surface brightness
into three dimensions, we adopted a Multi-Gaussian Expansion \citep[MGE; ][]{emsellem94}.
The MGE fit was performed with the method and software of \cite{cappellari02}, on the 
HST I-band images deconvolved from the PSF \citep[using a Tiny Tim PSF][]{krist95}.
As most of the clusters are barely resolved in the HST images and shape 
measurements are therefore impossible, we assume spherical symmetry. Note that 
although the NCs in NGC300 and NGC7793 (the best resolved) are indeed round, this may be 
due to their host disks being seen face-on. \citet{seth06} find that edge-on NCs can have 
quite disky outer isophotes.  

\begin{figure*}[tbp]
\begin{center}
\includegraphics[width=\hsize]{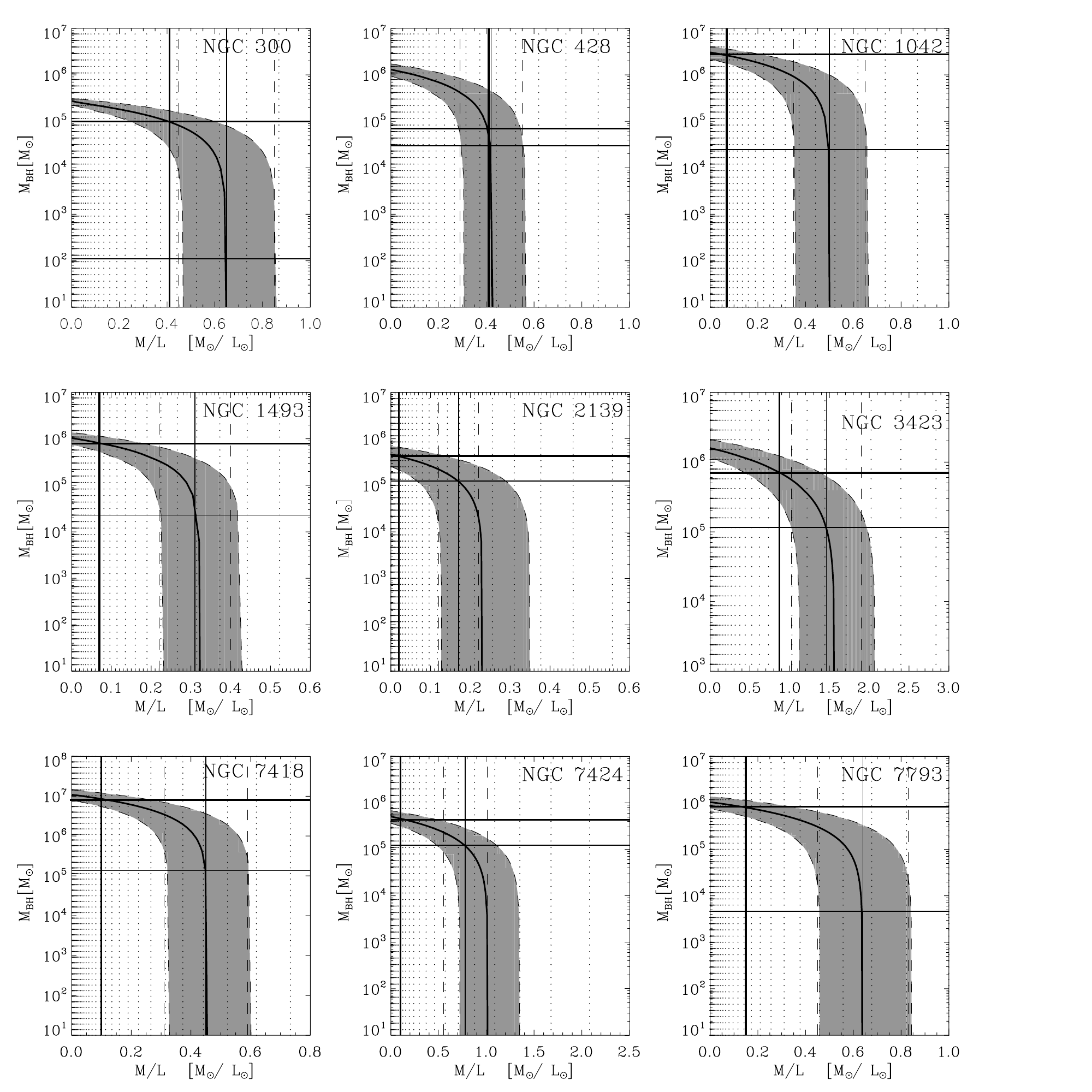}
\end{center}
\caption[grid]{The M/L ratio assumed for the stellar population against the mass 
of the putative black hole for each of the nine nuclear clusters. Models falling onto 
the right solid vertical line have the same velocity dispersion as given in Table 
\ref{t:values} for each cluster (dashed lines are upper and lower uncertainties). 
We also draw a vertical, full line (left) for the minimum 
mass-to-light ratio compatible with the observed spectrum of the stellar 
population in the cluster. The horizontal solid lines indicate the black hole masses 
referring to the two different M/L values quoted above. For the minimum M/L we find
a firm upper limit to the black hole mass M$_{BH}^{max}$ (upper line) and for the 
best-fit stellar populations M/L we get M$_{BH}^{best}$ (lower line).}
\label{f:grid}
\end{figure*}

We use the Jeans Anisotropic MGE (JAM) software by \citet{cappellari08} which implements
the solution of the Jeans equations allowing for orbital anisotropy. The model has three free 
parameters: (i) The anisotropy, (ii) the mass of a central black hole \mbh\ and (iii) the I-band 
total dynamical $M/L$. 
From the velocity dispersion profile computed by JAM, we compute the luminosity weighted 
velocity dispersion ($\sigma_{LW}$) over an aperture of 1 square arcsecond. This corresponds 
to the width of the UVES slit on the sky. We iterate the computation of $\sigma_{LW}$ over a grid of values 
for M/L and $M_{BH}$. The results are shown in Figure \ref{f:grid} which is directly comparable 
to Figure 8 of \citet{barth09}. Direct comparison with the mass-to-light ratios obtained by 
W05 (thin solid vertical line), shows that the ratios scatter around 1.0, with no obvious systematic outliers. The 
small differences in the result can be attributed to the way in which the surface brightness was modelled 
(Multi-Gaussian expansion here vs. direct deprojection in W05).

The maximum allowed mass of the black hole will be obtained when a minimum of mass is 
present in the form of stars. From Figure \ref{f:grid} one can easily read what BH mass 
would result if we  assumed M/L = 0 for the stars in the cluster. A more interesting lower 
limit to the M/L comes from the spectral fitting with stellar population models. We exploit the fact 
that the age 
obtained by fitting a simple single stellar population to a composite stellar population is strongly biased to the age of the 
youngest population in that object which contributes significantly to the total luminosity 
\citep[see e.g. W06, ][]{serra07}. The relevant values are tabulated in W06, and repeated 
in Table \ref{t:values} along with the values derived as upper limits to the mass of a putative 
BH from the intersection between both thick solid lines in Figure~\ref{f:grid}. This is a conservative estimate for the \mbh upper limit. A more realistic value for \mbh can be derived from the intersection or asymptotic point of the model (thick solid curved line) with the best-fit M/L from W05 (thin solid vertical line) in Figure~ \ref{f:grid}. The resulting best-fit \mbh values are listed in Table~\ref{t:values}.

We explicitly test the 
effect of velocity anisotropy on the modeling results and found very little change in the 
results - certainly below our systematic uncertainties due to the lower limit to the mass 
to light ratio that we apply \citep[see also][]{barth09}. We therefore neglect this effect for the rest 
of this paper.

\begin{table*}
\begin{center}
\caption{Properties of the sample of NCs in bulgeless galaxies}
\vspace{0.5cm}
\begin{tabular}{llcccccc}
Galaxy &  Type & NC r$_e$ & $\sigma$ & M/L$^{min}$ & M$_{BH}^{max}$ & M$_{BH}^{best}$  \\
            &           &   $(pc)$        &   $(km/s)$  &  $(M/L_{I,\sun})$ & {\it(\msun)}            &  {\it(\msun)} \\
\hline
NGC 300    & SAd & 2.9       & $13\pm2$      &  0.41  & 1$ \times 10^5 $ & 1$ \times 10^2 $  \\
NGC 428    & SABm & 3.36    &  24.4$\pm$4  &  0.41  & 7$ \times 10^4$ & 3$ \times 10^4 $  \\
NGC 1042  & SABcd & 1.94    & 32$\pm$5       &  0.07    & 3$ \times 10^6$ & 2.5$ \times 10^4  $ \\
NGC 1493  & SBcd &  2.6     &  25$\pm$4      &   0.07    & 8$ \times 10^5 $ & 2.5$ \times 10^5 $ \\
NGC 2139  &SABcd &  10.3   & 17$\pm$3       &  0.02   & 4$ \times 10^5 $ & 1.5$ \times 10^5 $ \\
NGC 3423  & SAcd &  4.18   &30$\pm$5        &  0.87    & 7$ \times 10^5 $ & 1.5$ \times 10^5 $ \\
NGC 7418  & SABcd & 12.3    &  34$\pm$5      &  0.10  & 9$ \times 10^6 $ & 1.5$ \times 10^5 $ \\
NGC 7424  & SABcd & 7.4      & 16$\pm$2       &  0.10     & 4$ \times 10^5 $ & 1.5$ \times 10^5 $ \\
NGC 7793  &  SAd &  7.7    & 25$\pm$4       &  0.15   & 8$ \times 10^5 $ & 5$ \times 10^3 $ \\
\end{tabular}
\label{t:values}
\end{center}
\end{table*}

\begin{table*}
\begin{center}
\caption{Sample of galaxies for which new properties were derived in this paper}
\vspace{0.5cm}
\begin{tabular}{llccccccl}
\hline
\hline
Galaxy &  Type & $\sigma$ & Dist     &  M$_{BH}$  & M$_{NC}$ & Sersic n & M$_{Bulge}$  & Ref \\
            &           &    $(km/s)$   & $(Mpc)$ &  {\it(\msun)}      &  {\it(\msun)}    &               &   {\it(\msun)}               &       \\
\hline
group 1 & & & & $\times$ & & $\times$& \\
\hline
NGC 300    & SAd      & $13\pm2$        & 2.2       & $<1 \times 10^2 $   & 1$ \times 10^6 $  & 1.1     & --    &  1 \\
NGC 428    & SABm   &  24.4$\pm$4    &  16.1        & $<3 \times 10^4 $   & 3$ \times 10^6 $  & 1.05   & --    & 1 \\
NGC 1042  & SABcd  & 32$\pm$5        &  18.2        & $<2.5 \times 10^4 $ & 3$ \times 10^6 $  & 1.15   & --    & 1 \\
NGC 1493  & SBcd    &  25$\pm$4       &  11.4       & $<2.5 \times 10^5 $ & 2$ \times 10^6 $   & 2.36	 & --	 & 1 \\
NGC 2139  &SABcd   & 17$\pm$3        & 23.6         & $<1.5 \times 10^5 $ & 8$ \times 10^5 $   & 1.53	 & --	 & 1 \\
NGC 3423  & SAcd    & 30$\pm$5        & 14.6        & $<1.5 \times 10^5 $ & 3$ \times 10^6 $   & 1.20	 & --	 & 1 \\
NGC 7418  & SABcd  &  34$\pm$5       & 18.4         & $<1.5 \times 10^5 $ & 6$ \times 10^7 $   & --	& --	& 1 \\
NGC 7424  & SABcd  & 16$\pm$2        & 10.9         & $<1.5 \times 10^5 $ & 1$ \times 10^6 $   & 0.91	 & --	 & 1 \\
NGC 7793  &  SAd     & 25$\pm$4        & 3.3        & $<5 \times 10^3 $   & 8$ \times 10^6 $  & 1.27   & --    & 1 \\
\hline
group 2 & & & & & $\times$ & & \\
\hline
NGC 4486  &  E1       & 375$ \pm18 $  & 17.0  & 6.3$ \times 10^9 $  &  $< 2 \times 10^8 $  &  6.86  & 6.0$ \times 10^{11}$ & 2 \\
NGC 4374  &  E1       & 296$ \pm14 $  & 17.0  &  1.5$ \times 10^9 $  & $<6.3 \times 10^7 $   &  5.60  & 3.6$ \times 10^{11}$ & 3 \\
NGC 1332  &  S0       & 321$\pm 14 $  &  19.6 & 1.45$ \times 10^9$ & $<1.4 \times 10^7$ &   --   & --        & 4 \\
NGC 3031  &  Sb       & 143$ \pm 7$    & 4.1    &   8$ \times 10^7$    & $<7 \times 10^6  $  &  3.26  & --        & 5 \\
NGC 4261  &  E2       & 315$ \pm15 $  &  33.4 &   5$ \times 10^8 $   & $<1.7 \times 10^6$ &  7.30  & 3.6$ \times 10^{11}$ & 6 \\
\hline
group 3 & & & & & $\times$ & & \\
\hline
NGC 4649   & E2      &  385$ \pm19 $  &  16.5 &  2.1$ \times 10^9 $  & $<2 \times 10^6 $  &  6.04  & 4.9$ \times 10^{11}$ & 7 \\
NGC 3998   & S0      &  305$ \pm15 $  & 14.9  & 2.4$ \times 10^8 $  & $<8.5 \times 10^5 $  &   --      & --        & 8 \\
NGC 2787   & SB0    & 189$ \pm9    $  &   7.9  & 0.7$ \times 10^8 $  & $<1.9 \times 10^6 $  & 1.97   &  --       & 9 \\
NGC 3379   & E0      &  206$ \pm10 $  & 11.7  & 1.2$ \times 10^8 $  & $<1.4 \times 10^4 $  &  4.29  & 6.8$ \times 10^{10}$ & 10 \\
NGC 4342   & S0      &  225$ \pm11 $  & 18.0  & 3.6$ \times 10^8 $  & $<2.5 \times 10^6 $  &  5.11  & 1.2$ \times 10^{10}$ & 11 \\
NGC 4291   & E2      &  242$ \pm12 $  &  25.0 & 3.2$ \times 10^8 $  & $<5 \times 10^6 $  &  4.02  & 1.3$ \times 10^{11}$ & 7 \\

\end{tabular}

\medskip

Note. --  Galaxies for group 1 are from W05, and we here derived upper limits on the black hole mass and Sersic~n. Galaxies for group 2 and 3 are from \cite{gultekin11}. For group 2 objects we derived upper limits on the NC mass via dynamical arguments, while for group 3 objects we used photometry to derive \mnc upper limits. For group 2 \& 3 Sersic n values are taken from \cite{graham07}, bulge masses are from \cite{haring04}, and velocity dispersions are from Hyperleda. The newly derived quantities are marked with an $\times$ at the top of the respective column.
\textsc{References for black hole masses}. - (1) this work, (2) \citealt{gebhardt09,gebhardt11}, (3) \citealt{bower98}, (4)
\citealt{rusli11}, (5) \citealt{devereux03}, (6) \citealt{ferrarese96}, (7) \citealt{gebhardt03}, (8)
\citealt{defrancesco06}, (9) \citealt{sarzi01}, (10) \citealt{gebhardt00b, shapiro06}, (11) \citet{cretton99}\label{t:values2}

\end{center}
\end{table*}

%RESULTS
\section{Global-to-nucleus relations}

We now plot the the upper limits we have derived into figures showing existing correlations 
from earlier work. In these figures we typically have a comparison sample which is taken from 
a larger statistical study and we add a number of objects at the low mass end from different 
sources in the literature. We have tried to be complete at the very lowest mass end of the 
relations. Further literature does exist, but typically, the BH masses exceed values of $\sim 10^6$\msun\ 
and the galaxies structural parameters have not been studied individually. 

\subsection{\mbh$-$ $\sigma$ relation}

For Figure \ref{f:msig}, the \mbh$-$ $\sigma$ relation, the comparison sample and relation are as compiled in 
\citet[][black open symbols]{gueltekin09}. We extend this compilation with recent 
maser measurements by \citet[][green crosses]{greene10}. 
Active AGN are denoted as blue stars; these are NGC4395 \citep{filippenko03, peterson05} and POX52 
\citep{barth04}. In principle, NGC1042 from the present work falls also into this category \citep[see][]{shields08}, 
but is plotted as a filled red circle. Previous upper limits for non-active nucleated galaxies are plotted as open blue circles: 
M33 \citep{gebhardt01, merritt01}, NGC205 \citep{valluri05}, IC342 \citep{boker99}, and NGC404 \citep{seth10}. 
We also plot the globular clusters G1 \citep{gebhardt05}, $\omega$ Cen \citep{noyola08, noyola10, jalali11}, and
NGC6388 \citep{luetzgendorf11} as green open circles. The verdict on the usefulness of these measurements is still 
out, with strong contrasting claims by other authors that there is no evidence for a black hole in $\omega$ Cen 
\citep{anderson10,marel10} and in G1 \citep[e.g.][]{baumgardt03}. 
We nevertheless use the derived values in a spirit of adventure, i.e. what would 
it mean if these measurements were correct? Finally, the new upper limits 
derived in this work are denoted as filled red circles. It emerges that a flattening of the relation is not 
consistent with the current measurements. It may well be that a downwards bending would be necessary, if 
more stringent upper limits such as that for M33 would be published. 

\subsection{\mbh$-$ \mbulge relation}

For Figure \ref{f:sph_BH}, the \mbh $-$ \mbulge relation, the comparison relation and sample are taken 
from \citet{haring04} (filled black circles), while the other data points come from the same sources as in 
Figure \ref{f:msig}. There is a hint towards a flattening of the relation with the lowest spheroid masses, but 
it will be difficult to confirm this without much better estimates of the masses of IMBHs. On the other hand a 
steepening, i.e. bulges that are too massive for their BHs, has been mentioned by \citet{greene08, greene10}. 
If there are BHs in galaxies with no bulges as well as bulges that are too massive for their BHs, it seems clear that 
the \mbh $-$ \mbulge relation must suffer from large scatter at small masses. 

\subsection{\mbh$-$ \nsersic relation}

\begin{figure}[tbp]
\begin{center}
\resizebox{0.99\hsize}{!}{\includegraphics[]{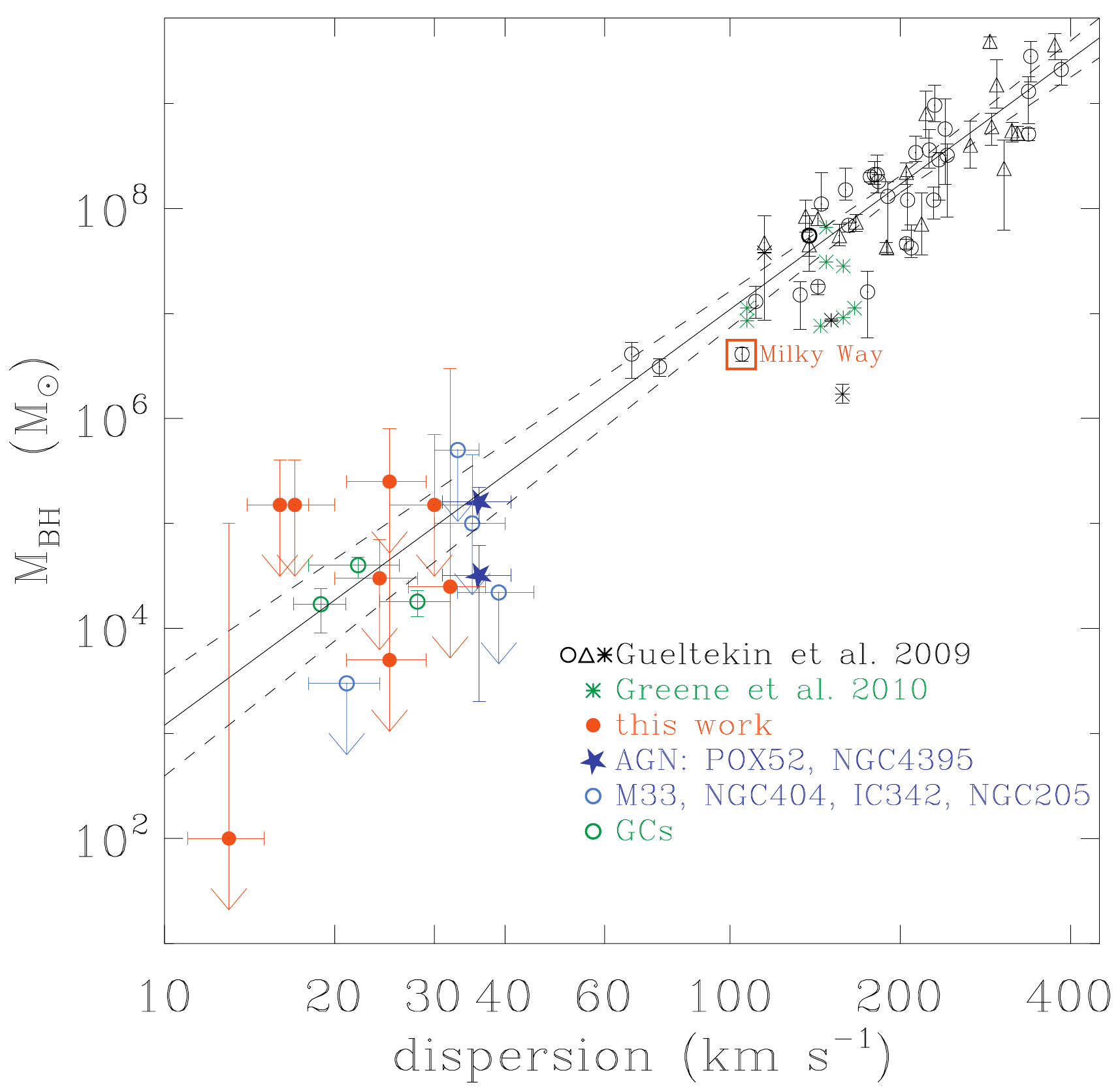}}
\end{center}
\caption[mbhsig]{The relation between the mass of the BH and the velocity dispersion 
of the spheroid around it. We plot the objects as listed in the text. The lines give the best fit of \cite{gueltekin09}. }
\label{f:msig}
\end{figure}

\begin{figure}[tbp]
\begin{center}
\resizebox{0.99\hsize}{!}{\includegraphics[]{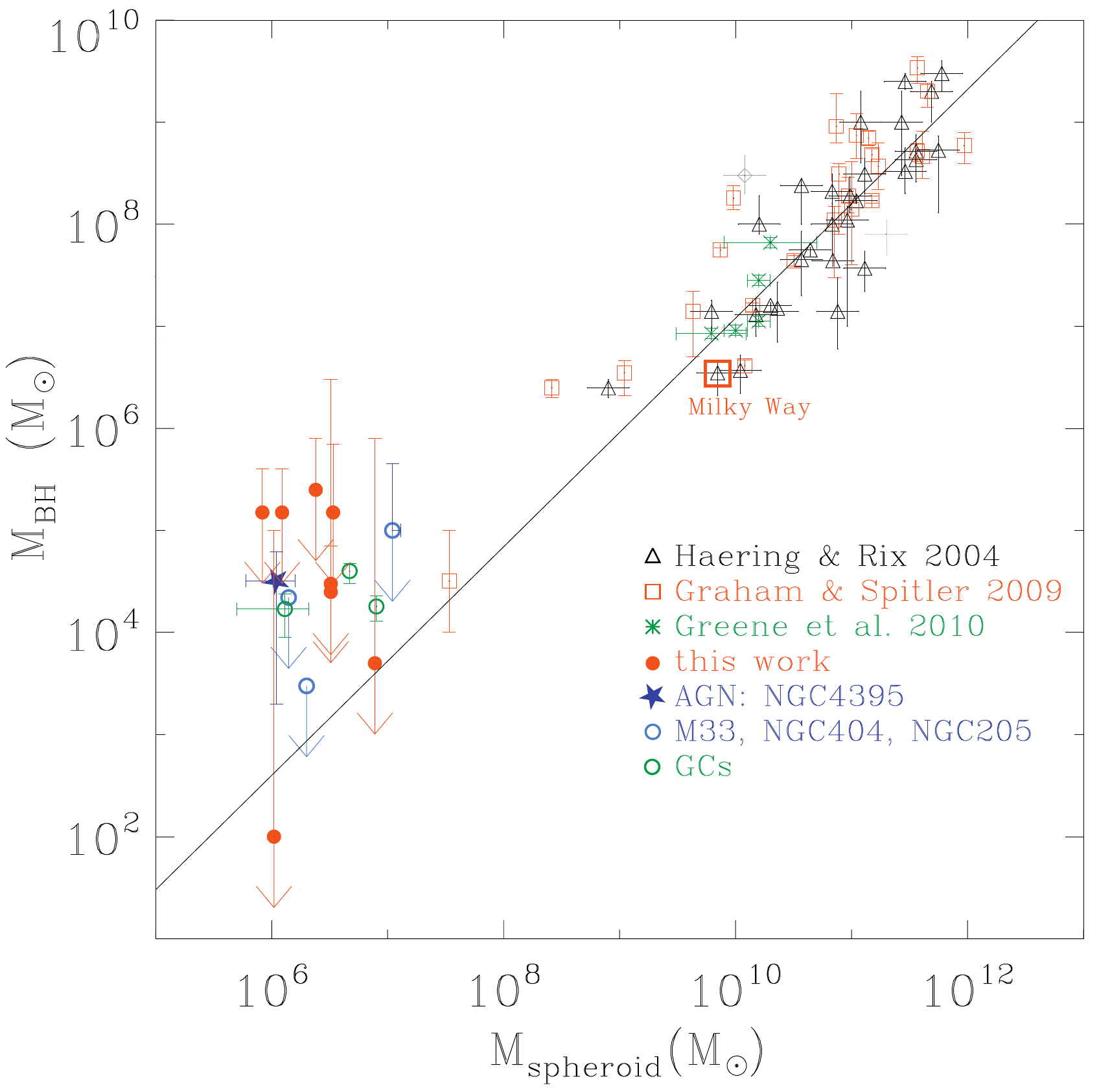}}
\end{center}
\caption[mbhsig]{The BH mass vs. the spheroid mass (bulge, GC, NC). 
We plot the objects as listed in the text. The line indicates the best-fitting relation of \cite{haring04}}
\label{f:sph_BH}
\end{figure}

\begin{figure}[tbp]
\begin{center}
\resizebox{0.99\hsize}{!}{\includegraphics[]{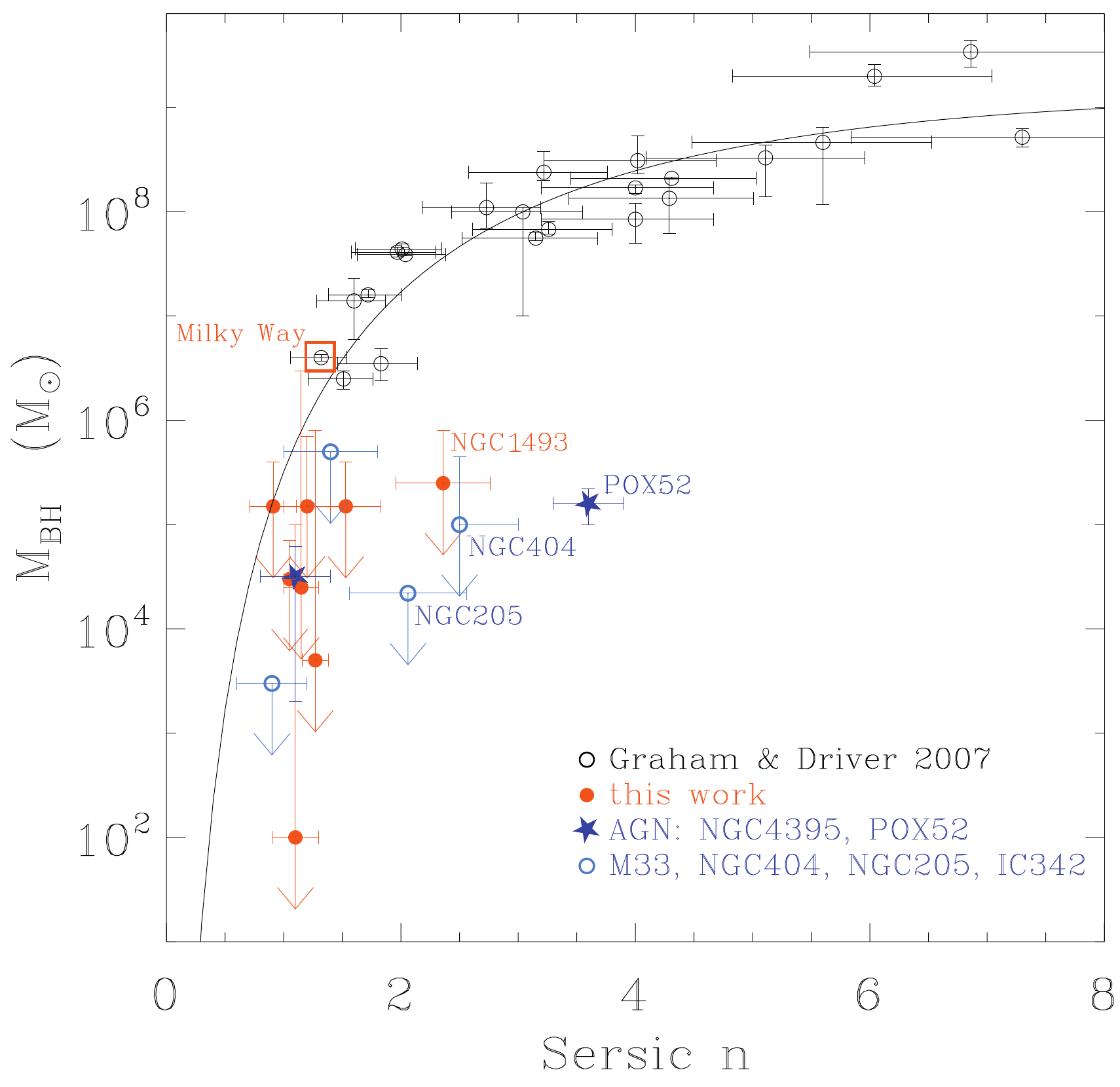}}
\end{center}
\caption[mbhsig]{The mass of the BH against the Sersic index of the host bulge or disk. 
We plot the objects as listed in the text. The largest outliers are NGC205 (with n=2.05), POX52 (n=3.6), NGC404 (n=2.5), and NGC1493 (n=2.4).}
\label{f:sersic}
\end{figure}

For Figure \ref{f:sersic}, the \mbh $-$ \nsersic relation, the comparison relation and sample are taken 
from \citet{graham07} (filled black circles), while the other data points come from the same sources as in 
Figure \ref{f:msig}. We have also assembled measurements of the Sersic n from literature sources for all 
objects with published BH masses. For the galaxies with \mbh limits newly derived in the present paper, 
Sersic n was derived from the following literature sources: \citet{ganda09} for NGC1042 and NGC3423, 
\citet{weinzirl09} for NGC2139. For NGC300 and NGC428 Spitzer IRAC 3.6$\mu$m images were 
downloaded from the Spitzer Heritage Archive\footnote{http://sha.ipac.caltech.edu/} 
and Sersic n was determined using the GALFIT software \citep{peng02}. For NGC~205 we used the surface brightness profile
of \citep{valluri05} to derive Sersic n. For NGC1493, NGC7424 and 
NGC7793 the corresponding images were obtained from the 2MASS archive\footnote{http://irsa.ipac.caltech.edu/} 
and again fitted with GALFIT. All galaxies were fit using one PSF component for the central NC, one Sersic component 
representing the disk and one constant sky component. All parameters were left free to be 
fit for. The webpages provide appropriate point spread functions, although all 
of our targets are nearby and therefore well resolved, the resulting Sersic n is almost independent of 
the PSF used in GALFIT. We caution that the resulting Sersic n may depend heavily on the radial range used in the 
fitting. To cite two extreme examples, the Sersic n of NGC300 is independent of the radial range used 
within $\Delta(n) = 0.1$. On the other extreme, the Sersic n for NGC1493 varies between $\sim 1.3$ and the reported 
value of $\sim 2.5$. It is beyond the scope of the current paper to derive a physically meaningful fit 
range that would put the physical meaning of the Sersic n on firmer ground. We emphasize that it is 
\emph{despite} the cited uncertainties that the relation between \mbh and Sersic n seems to hold. 

Figure \ref{f:sersic}, shows two interesting features: 1)~Because the relation 
fitted by \citet{graham07} curves down at $n$=1, a large range of BH masses is allowed in this regime, 
which clearly allows for the scatter that seems to emerge as a common trend in the previous two nucleus-to-global 
relations. 2) There are significant outliers in this plot, in the sense that some low-mass galaxies can have too high 
Sersic $n$ for their BH mass. 

\subsection{\mbh$-$ \mnc relation}

\begin{figure}[tbp]
\begin{center}
\resizebox{0.99\hsize}{!}{\includegraphics[]{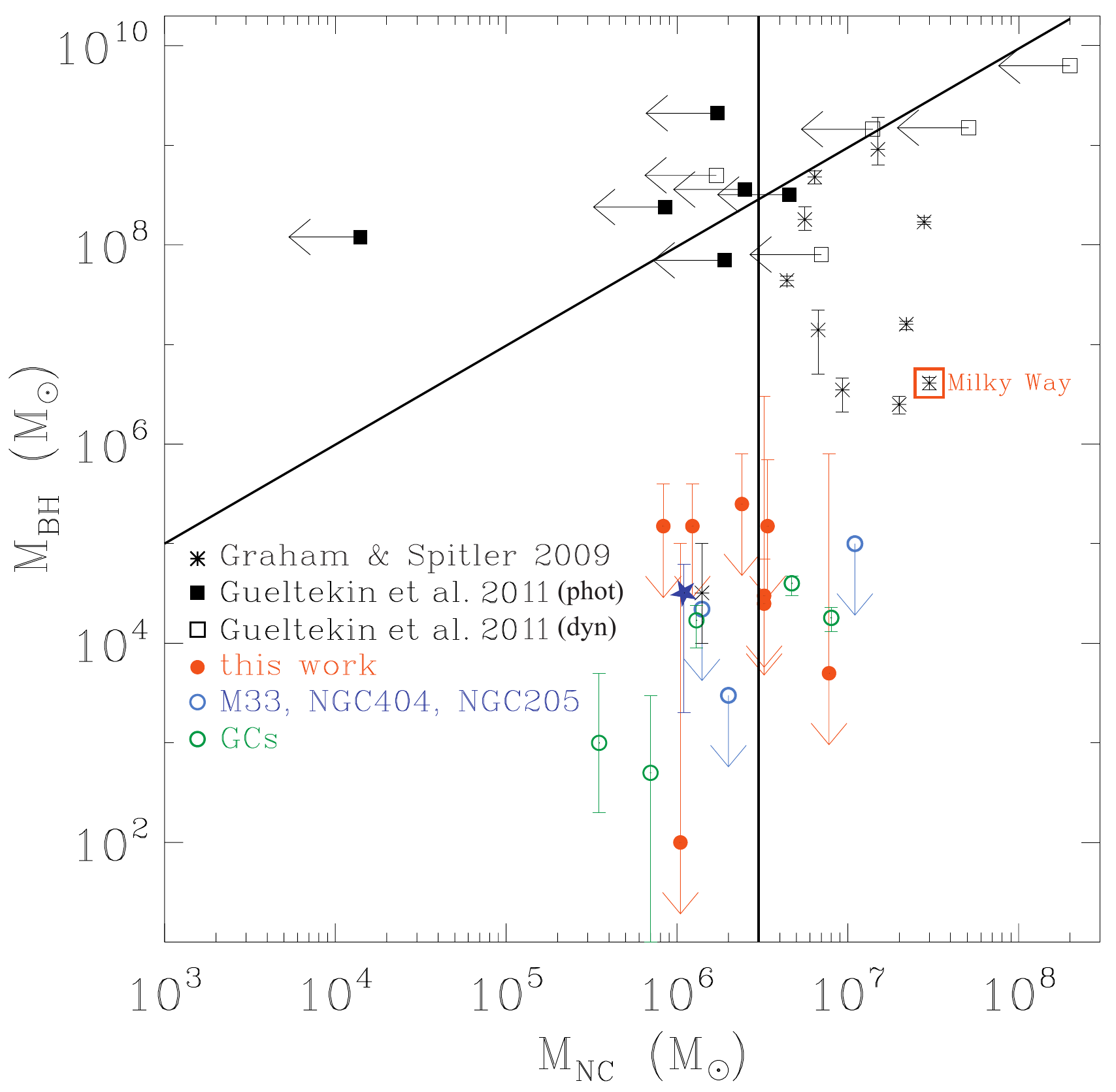}}
\end{center}
\caption[mbhsig]{The mass of the BH mass vs. the NC mass. 
We plot the objects as listed in the text. The two full lines indicate 
a NC mass of $3 \times 10^6$ \msun and a \mbh / \mnc mass ratio 
of 100. These lines separate NC dominated galaxy nuclei (lower left 
of both lines) from BH dominated galaxy nuclei (upper left of both lines) 
and a transition region (to the right of both lines). 
}
\label{f:NC_BH}
\end{figure}

In Figure \ref{f:NC_BH} we show the relation between \mbh and \mnc \citep[compare][]{graham09b, seth08}. 
We have plotted objects already used above, for which determinations of both \mbh and \mnc  exist.  In searching for a high 
mass comparison sample we have made use of the compilations by \citet{graham09b} and \citet{gultekin11} from which we also 
take the distances. Where not available, 
we have then proceeded to derive upper limits to the NC masses either from the literature or from own fits to 
archival HST images\,\footnote{Thorough work deriving consistent photometry and structural parameters for NCs across the 
entire Hubble sequence for large swaths of the HST archive is badly needed, but is beyond the scope of the current work. 
Note that one focus of such work could be the distinction (if any clear distinction exists) between NCs and nuclear 
disks. In the case of NGC4342, for example, the upper limit we give on the NC mass is not only observationally uncertain, but 
also conceptually uncertain. As \cite{scorza98} discuss, a relation between the nuclear disk mass and the BH is as plausible 
as between the NC and the BH. Indeed, some NCs may turn out to be nuclear discs on close inspection \citep[compare][]{seth06}.}.

We now discuss the ways that we have obtained upper limits for the NC masses galaxy by galaxy. We strongly emphasize that 
we have tried to obtain \emph{upper limits to} rather than real measurements of the NC mass. Real measurements of NC masses 
can only be carried out by a combination of 
dynamical modeling and spectral analysis to determined the relative influence of the AGN and possible varying M/L ratios. 
We rather aim to be conservative with respect to all uncertainties affecting our estimates of upper limits to the NC masses. Our resulting upper limits are listed in Table~\ref{t:values2}.
For the following 5 objects we estimated upper mass limits from the literature only. 

NGC4486 (M87): the bright nucleus is dominated by AGN light. There is no evidence for a NC. We therefore use Figure 7 
of \cite{gebhardt09}, which shows the enclosed stellar mass within the central arcsec to be $2 \times 10^8$ \msun. 
This is consistent with an estimate from \cite{young78}, which gives a total of M = $5 \times 10^9$ \msun\ within a radius 
of 100pc and M/L=60, thus leading to an estimate of the stellar mass within that radius 
of $3\times 10^8$ \msun, assuming that the stellar M/L = 4. We emphasize that this is the total stellar mass within a radius 
comparable to the radii of typical NCs, and therefore gives an upper limit to \mnc. We do not claim that M87 actually 
hosts a stable NC at it centre.

NGC4374 (M84): an AGN has been shown to exist by \cite{bower00}, with very weak stellar features. To estimate an upper mass to the NC in NGC4374 we use the paper by \citet{walsh10}, which yields a BH mass estimate of $4\times10^8$ \msun. Their Figure 4 shows the circular velocity profiles due to the BH and the stellar mass, respectively. Assuming a distance to M84 of 17 Mpc, yields 70 pc / arcsec. Assuming a NC radius of 10\arcsec and a stellar M/L=4, we obtain that at 10 pc radius the circular velocity due to the BH is 400 km/s, while the circular velocity due to the remaining stellar mass is smaller than 50 km/s. To obtain an estimate of the upper limit for a putative NC, we need to correct for the different masses and for the different spatial distribution (point-like vs. extended). From the virial theorem, we can scale the velocity quadratically. From Table 2 in the current work it can be seen that a conservative factor (i.e. one that gives a lot of stellar mass) for the conversion from point-like to extended would be a factor of 10. The upper limit for the stellar mass within 10 \arcsec then becomes: \mnc / \mbh = $10 * 50^2 / 400^2$ = 0.15, thus yielding a NC upper mass limit of $6.3\times10^7$\msun. \citet{walsh10} also state that 
stellar mass is a negligible contributor to their mass budget, it is therefore entirely possible that no NC exists in that galaxy. 

NGC4261: The central luminosity distribution is complex, with a nuclear disk and a luminous nuclear source which seems to be dominated by  
an AGN, at least a radio jet is present \citep{ferrarese96}. There is thus no clear evidence in favour of any NC.
\cite{ferrarese96} find that M/L$_V$ = 2100 within the inner 14.5 pc. A maximum M/L$_V$ for stellar populations is 7. We thus obtain that 7 / 2100 of the 
central mass within 15 pc can be in stars, which is $5 \times 10^8$ / 300 = 1.6 $\times 10^6$ \msun. 

NGC1332: There is no firm evidence for a NC, although the surface brightness profile of  \cite{rusli11} hints at a central luminosity excess 
within the central arcsec. The dynamical model of \cite{rusli11} gives a central stellar luminosity density of $4 \times 
10^{12}$ L$_{\odot}$ kpc$^{-3}$. 
For a NC of 5 pc radius this yields a NC luminosity of 2 $\times 10^6$ L$_{\odot}$ in the R-band. With M/L$_R =$ 7 
\citep[also according to][]{rusli11}, \mnc max is 1.4 $\times 10^7$ \msun. 

NGC3031: \cite{devereux03} list values of stellar mass within radius in their Table 3. From their Figure 3, it is clear that the nuclear source is not extended, there is thus no evidence for the presence of a NC. To estimate an upper mass limit for the NC, we assume a NC radius of 7pc \citep[compare][]{boker02,cote06}, the upper limit to \mnc is then 7 $\times 10^6$ \msun.

For the following 6 objects no NC mass estimate was available. We therefore turned to the HST images as downloaded from the Hubble 
Legacy Archive. We have then used GALFIT \citep{peng02} to derive the magnitudes of the NCs. Because all NCs 
we treat in this last step are in early-type galaxies, we can assume that their ages range between 1 and 10 Gyr, yielding 
an estimate of the allowed range for the M/L ratio. For most cases we used the F814W filter on either ACS or WFPC2, 
setting the allowed range of M/L between 1 and 4. Much more sophisticated modeling of the photometry, while possible, 
would yield only marginally better estimates of the total stellar mass of the NC for several reasons: 1) The star formation 
histories (SFHs) of NCs are unknown and indeed, likely to be semi-random, repetitive bursts of star formation. Therefore 
no strong prior can be applied to the SFH. Because the oldest stellar populations are the faintest per unit mass, the 
resulting uncertainty on M/L is of order a factor 2. 2) The photometry of the NCs is in itself uncertain. We have made 
use of realistic PSFs from either \citet{jee07}\footnote{http://acs.pha.jhu.edu/\~mkjee/acs\_psf/} or 
from Tiny Tim\footnote{http://www.stsci.edu/hst/observatory/focus/TinyTim}. It is much less certain, 
what the ideal profile 
for the surface brightness of the host galaxy should be though \citep[compare][]{ferrarese06b,lauer07}. We have used one single Sersic, 
as we are only interested in subtracting the host, not in describing it. Nevertheless, we estimate that the use of different 
profiles (2 Sersics, Nuker, etc.) could impact the total photometry of the NC by up to 0.5 or even 1 magnitude 
\citep[compare e.g. the central extrapolations of][]{boker02}. We therefore 
have chosen to let these uncertainties be reflected in the errorbars of the NC mass estimate, rather than trying to hide them 
somewhere within a sophisticated analysis. 

NGC4649: No nuclear source is visible \citep[as also found by ][]{graham09b}. We first fit this galaxy with a single Sersic. When additionally forcing in a point source (GALFIT PSF component) of different magnitudes (20, 20.5, 21,21.5, 22), the resulting oversubtraction can be seen clearly in the residual image for as faint as m$_I$ = 21.0. We use this value as a conservative upper limit to the NC magnitude. This results in an upper mass limit of $ 2 \times 10^{6} $ \msun.

NGC4291: We attempted the same procedure as before. However, due to a flat central surface brightness profile, our simple Sersic fit by itself 
produced an oversubtraction of the central flux, not allowing us to use the exact same procedure as for NGC4649.  Nevertheless, 
the HST image clearly shows the absence of any point source in the center. We therefore assumed the same limit as before, i.e. 21 mag in F814W, which results in \mnc$ = 5 \times 10^{6} $ \msun.

NGC3998: After the GALFIT fit, a clear spiral structure and a bar are seen in the residuals. The central light source was modeled as a Sersic with an effective radius of 0.2\arcsec\ and a Sersic n = 0.1, making us believe it is unresolved. \citet{defrancesco06} classify this galaxy as a LINER, thus the central source is AGN-light dominated. Therefore our photometrically derived NC mass of $ 8.5 \times 10^{5} $ \msun again is a conservative upper limit.  

NGC4342: The fit with GALFIT was difficult, with 4 Sersic components in the final fit. The final solution was chosen to oversubtract the 
NC. Again we have a conservative upper limit of 21.85 mag corresponding to \mnc$= 2.5 \times 10^{6} $ \msun, using an M/L of 6.5 in I from \citet{cretton99}. Contamination from AGN light is also possible, making our upper mass limit more robust.

NGC3379 (M105): The NC is visible in \citet{gebhardt00b}, but not mentioned there. \citet{graham09b} note this galaxy as un-nucleated. 
Two extended components with Sersic n $\sim 1$ and one very compact source with Sersic n $\approx$ 1/2 (i.e. Gaussian surface brightness profile) and 
r$_e$ = 0.2 \arcsec\ give a good fit to this object. The measured integrated magnitude of the central point source is 25.7 in the F814W 
of WFPC2, corresponding to \mnc$ = 1.4 \times 10^{4} $ \msun. We used M/L$_F814W$ = 3 as a suitable upper limit to the M/L.

NGC2787: This galaxy was analyzed in \citet{peng02} and the nuclear photometry is taken from that source. We used M/L$_F547M$ = 3 as a suitable upper limit to the M/L. Thus, the NC upper mass limit is $ 1.9 \times 10^{6} $ \msun.

Note that in the galaxies NGC4486, NGC4374, and NGC3379 a luminous nuclear source is clearly seen. While this could all 
be AGN light, we see no way to ascertain the absence of a NC. In contrast to \citet{graham09b} we
only claim to be able to derive an upper limit to the NC mass, rather than excluding a NC alltogether. 
Note also that a stellar cusp containing 10\% of the BH mass is predicted around any BH \citep{merritt06a}.

\section{Discussion}

We now discuss and interpret a number of features we saw in the previous section, with the aim to discuss 
ideas that emerge from these Figures but to our knowledge, have not been discussed
 in the literature before. The ultimate aim of our 
study is of course to contribute to a consistent physical picture of black hole and nuclear cluster growth. 

In Figure \ref{f:sersic}, one relation between the global galaxy properties and \mbh holds for a large range of 
$n$, independent of the presence of a bulge. Outliers are rather low-mass galaxies (and not low-mass black holes). 

We stress that this relation is purely observational at this stage. Due to the heterogeneous assembly of the Sersic n 
values (literature, own fitting) the physical region represented by them is not always the same. In particular some 
of the galaxies do not contain a bulge, while for others the Sersic n has been explicitly measured for the bulge 
component. The existence of a relation seems evidence that indeed the measurement of Sersic n is meaningful. 
In particular, no conspiracy is obvious that would fundamentally bias our measurements in the sense of producing 
 a spurious correlation. It thus seems to us that even independently of the exact details of the derivation of the 
 Sersic n, it clearly describes a property of the galaxy that is relevant for the BH mass. A question to ask 
is then, whether we fully understand the physical implications of that relation, and whether we could 
potentially reduce the relation to underlying intrinsic distribution of galaxy properties (e.g. if Sersic n was 
related to bulge mass in a very tight manner, we might be tempted to argue that bulge mass is the more 
fundamental measurement). We believe the present paper cannot resolve this question but hope it provides 
motivation to explore these issues further. 

To venture a possible physical interpretation of the outliers from the relation we note the following: it could be 
that the transformation process from disk galaxy to spheroid 
is different in this galaxy mass regime. While BHs in massive galaxies grow during the morphological transformation 
process of their host galaxies, BHs in low mass galaxies are not affected (fed) during the transformation 
process. It might be worthwhile exploring through simulations, whether this has to do with a possible transformation 
dichotomy, i.e. mergers vs. harassment. It is worth pointing out here that such a dichotomy does not seem to 
be immediately apparent from the age or metallicity profiles, as these seem not to depend on mass \citep{koleva11}. 

Figure \ref{f:NC_BH}  has not been published previously in this form to the best of our knowledge 
\citep[though compare][for similar representations]{graham09b, seth08} and 
may yield considerable insight on the relation 
between NCs and BHs. An immediate conclusion from this figure is that BHs and their host NCs do 
not share the same intimate connection as BHs and their host spheroids. It rather seems that in galaxies with a 
high total mass, or alternatively a sizable spheroid, the BH has been able to grow \emph{independently} of the NC, 
thus being able to reach comparable masses. In galaxies or star clusters unaffected by spheroid growth, 
as e.g. the GCs, M33 and others, it seems the BH, if existent, is only a very small portion of the mass
of the NC. 

Figure \ref{f:NC_BH} \citep[and similar Figures, see][]{seth08,graham09b,graham11} is still in an early phase and we 
believe further studies in the field will attempt to fill in the high 
and low mass end of the BH mass regime with more NC masses and BH masses, respectively. Nevertheless it 
seems that two extreme ends can be identified, with a transition region in between. At the low BH-mass end, 
there is little evidence for the presence of any BH, yet NCs seem to be common (B02). On the other hand no nuclear BH has been 
found that is not surrounded by a NC in this regime. If GCs with BHs are indeed the remnants of accreted satellite galaxies 
\citep[e.g.][]{zinnecker88,freeman93,bekki03,boker08} and if indeed they lie on the  \mbh$-$ $\sigma$ relation, this 
would imply however, that at this stage BH growth is linked to NC growth much in the same way that BH growth is 
later tied to galaxy growth. A close look suggests indeed that some physical processes that occur in or with NCs, such 
as gas accretion \citep{milosavljevic04, bekki06, pflamm09} and merging \citep{tremaine75,capuzzo93,mclaughlin95, 
bekki04, miocchi06, agarwal11}, are quite similar to those experienced by galaxies. An alternative to the assumption that 
the process giving rise to the \mbh$-$ $\sigma$ is so astoundingly generic is of course that the BH mass measurements in GCs 
may be subject to the "expectation bias", i.e. when the measurement is in accordance with the expectations of the community 
they get accepted more easily. 

At the very high mass end of the BH mass range, the BH is much more massive than the NC. On the other hand, this is the 
region where global-to-nucleus relations hold best. This could happen through two mechanisms: 1) either the galaxies in 
question never had a sizable NC, possibly because their central BHs grew early on in the age of the universe, thus stopping 
NC growth \citep{nayakshin09}. Or 2) massive BHs destroy their host NCs. Figure \ref{f:NC_BH} in its current form suggests 
that this may happen at a mass ratio of $\ge$ 100 or alternatively when the BH radius of influence is of the same size as the NC 
radius. Loss cone depletion and core formation in early-type galaxies are well-studied 
mechanisms, that would amply suffice to destroy the pre-existing NC \citep{merritt06b}. 

\citet{bekki10}  have examined an alternative solution for the disappearance of NCs for massive galaxies. Their argument relies 
on the mergers that are responsible for the morphological transformation from disk-dominated to bulge-dominated galaxies. 
They show that NCs can be significantly heated and thus be made susceptible to destruction during the merger event. 
The picture painted here differs significantly from that painted in \citet{bekki10} in that we put weight on the importance of 
the BH for destroying the NC. Indeed, what determines NC disappearance does not seem to be galaxy morphology, 
as most early-type galaxies have NCs. Rather, there is evidence for an upper limit to the BH/NC mass ratio, arguing strongly for a pivotal 
role of this ratio in leading to NC disruption. 

The intermediate mass or transition regime may possibly lie between two boundaries, i.e. above NC masses of 5 $\times 10^6$ 
M$_{\odot}$ and below a \mbh / \mnc mass ratio of 100. In this intermediate mass regime, while BHs have grown by at least 2 orders 
of magnitude, and probably more than 4 as compared to the NC-dominated regime, the NC grows by at most a factor of 10. There thus 
is some common growth, yet it does not occur in parallel. On the other hand, this is the region of most scatter in the typical global-to-nucleus 
relations. This phase would thus be characterized as a transition phase between NC-dominated nuclei and BH-dominated nuclei. 

Does Figure \ref{f:NC_BH} imply that NCs do not grow by the same processes as their BHs and is this a serious setback
to the grouping together of NCs and BHs into CMOs \citep{ferrarese06a}? That NCs and BHs need not grow in parallel
has been emphasized by \cite{nayakshin09}, where both types of objects rather grow in competition for the same gas 
reservoir. \cite{nayakshin09} ask whether the BH can prevent the NC from growing through its feedback, 
and postulate that this is the case when the gas accretion rate is smaller than the Eddington rate. This picture is attractive 
in explaining Figure \ref{f:NC_BH} because it naturally explains the three regimes -- NC dominated, NC/BH transition, 
BH dominated. Nevertheless, given the very low accretion rates observed in bulgeless galaxies and the presence of 
significant BHs in at least a few of them, this picture seems to break down exactly for the NC-dominated regime. 

Discrimination between the different scenarios envisaged in the literature seems to be mostly an observational question at present. 
At low masses the error bars on 
BH measurements are typically very large, while NC masses are well measured. At high masses, BH masses are more 
accurate while the uncertainties for NC masses increase, due to resolution problems of the NCs above the underlying galaxies.
We need both reliable BH and NC masses to see what the exact locus of points in this plot is. If there is a smooth transition, 
making the sequence look like a closed parenthesis, this would imply that the destruction of the NC due to the growing black 
hole is a slow process. If there really is a well-defined transition at \mbh / \mnc = 100 then this would imply either that the process 
of NC destruction is very fast or that these galaxies never had a NC.

\section{Conclusions}

We have computed new upper limits for the masses of intermediate mass black holes 
in 9 pure disk galaxies with very low BH masses. We also computed upper limits to the 
masses of nuclear star clusters in the nuclei of galaxies with previously determined massive BHs. 
We plot these upper limits on the three global-to-nucleus relations \mbh - $\sigma$, \mbh - \mbulge 
and \mbh - Sersic n, as well as on a new Figure that compares \mbh and \mnc. We discuss the 
features we see in these figures. Two possible conclusions emerge from our discussion: 

\begin{enumerate}
\item In the \mbh - Sersic n figure, those galaxies that lie on the relation seem to prove that there is 
a relation between \mbh and the morphological transformation of their host galaxies. A few notable 
outliers are dwarf elliptical galaxies, where the morphological transformation process does not seem 
to be associated with BH growth. We speculate that this difference may arise from different mechanisms, 
i.e.~mergers for high mass galaxies and harassment for dwarfs. 
\item In the \mbh - \mnc figure, we can clearly distinguish three regimes, NC dominated, BH dominated and 
transition between the two. We speculate that this could imply that BHs are formed in NCs, then start to grow 
much faster than their host NCs and, through a transition phase with similar masses for both components, 
could then ultimately destroy their host through loss cone depletion. 
\end{enumerate}

We expect further progress in the field to arise from better measurements of BH masses at the low mass 
end of the \mbh mass function and from better measurements of NC masses at the high mass end of 
the \mbh mass function. In particular, it might be useful for further research in the field if authors attempting 
to measure black hole masses also stated more clearly what their constraints on the NC mass are. Currently 
NCs are treated more or less as a nuisance to get rid of, while a clearer assessment of the constraint on their 
mass would benefit our understanding of the role NCs play in galaxy nuclei.

\acknowledgments

We thank the referee for an extensive report that significantly improved the presentation of the results in this 
paper, in particular on the NC upper limits. 
The authors acknowledge the support and hospitality of the ESO Garching office 
for science during the genesis of this work. NN acknowledges support by the DFG cluster of excellence `Origin and Structure of the Universe'.
This research has made use of the NASA/IPAC Extragalactic Database (NED) which is operated by the 
Jet Propulsion Laboratory, California Institute of Technology, under contract with the National Aeronautics and Space Administration. 
We acknowledge the usage of the HyperLeda database (http://leda.univ-lyon1.fr). This work is based in part on observations made with the NASA/ESA Hubble Space Telescope, and obtained from the Hubble Legacy Archive, which is a collaboration between the Space Telescope Science Institute (STScI/NASA), the Space Telescope European Coordinating Facility (ST-ECF/ESA) and the Canadian Astronomy Data Centre (CADC/NRC/CSA).This work is based in part on observations made with the Spitzer Space Telescope, which is operated by the Jet Propulsion Laboratory, California Institute of Technology under a contract with NASA. This publication makes use of data products from the Two Micron All Sky Survey, which is a joint project of the University of Massachusetts and the Infrared Processing and Analysis Center/California Institute of Technology, funded by the National Aeronautics and Space Administration and the National Science Foundation. This research has made use of 
NASA's Astrophysics Data System Bibliographic Services.

\bibliographystyle{apj}
\bibliography{NCBH_biblio}

\begin{thebibliography}{113}
\expandafter\ifx\csname natexlab\endcsname\relax\def\natexlab#1{#1}\fi

\bibitem[{{Agarwal} \& {Milosavljevi{\'c}}(2011)}]{agarwal11}
{Agarwal}, M. \& {Milosavljevi{\'c}}, M. 2011, \apj, 729, 35

\bibitem[{{Anderson} \& {van der Marel}(2010)}]{anderson10}
{Anderson}, J. \& {van der Marel}, R.~P. 2010, \apj, 710, 1032

\bibitem[{{Barth} {et~al.}(2008){Barth}, {Greene}, \& {Ho}}]{barth08}
{Barth}, A.~J., {Greene}, J.~E., \& {Ho}, L.~C. 2008, \aj, 136, 1179

\bibitem[{{Barth} {et~al.}(2004){Barth}, {Ho}, {Rutledge}, \&
  {Sargent}}]{barth04}
{Barth}, A.~J., {Ho}, L.~C., {Rutledge}, R.~E., \& {Sargent}, W.~L.~W. 2004,
  \apj, 607, 90

\bibitem[{{Barth} {et~al.}(2009){Barth}, {Strigari}, {Bentz}, {Greene}, \&
  {Ho}}]{barth09}
{Barth}, A.~J., {Strigari}, L.~E., {Bentz}, M.~C., {Greene}, J.~E., \& {Ho},
  L.~C. 2009, \apj, 690, 1031

\bibitem[{{Baumgardt} {et~al.}(2003){Baumgardt}, {Makino}, {Hut}, {McMillan},
  \& {Portegies Zwart}}]{baumgardt03}
{Baumgardt}, H., {Makino}, J., {Hut}, P., {McMillan}, S., \& {Portegies Zwart},
  S. 2003, \apjl, 589, L25

\bibitem[{{Bekki} {et~al.}(2004){Bekki}, {Couch}, {Drinkwater}, \&
  {Shioya}}]{bekki04}
{Bekki}, K., {Couch}, W.~J., {Drinkwater}, M.~J., \& {Shioya}, Y. 2004, \apjl,
  610, L13

\bibitem[{{Bekki} {et~al.}(2006){Bekki}, {Couch}, \& {Shioya}}]{bekki06}
{Bekki}, K., {Couch}, W.~J., \& {Shioya}, Y. 2006, \apjl, 642, L133

\bibitem[{{Bekki} \& {Freeman}(2003)}]{bekki03}
{Bekki}, K. \& {Freeman}, K.~C. 2003, \mnras, 346, L11

\bibitem[{{Bekki} \& {Graham}(2010)}]{bekki10}
{Bekki}, K. \& {Graham}, A.~W. 2010, \apjl, 714, L313

\bibitem[{{Binggeli} {et~al.}(2000){Binggeli}, {Barazza}, \&
  {Jerjen}}]{binggeli00}
{Binggeli}, B., {Barazza}, F., \& {Jerjen}, H. 2000, \aap, 359, 447

\bibitem[{{B{\"o}ker}(2008)}]{boker08}
{B{\"o}ker}, T. 2008, \apjl, 672, L111

\bibitem[{{B{\"o}ker} {et~al.}(2002){B{\"o}ker}, {Laine}, {van der Marel},
  {Sarzi}, {Rix}, {Ho}, \& {Shields}}]{boker02}
{B{\"o}ker}, T., {Laine}, S., {van der Marel}, R.~P., {Sarzi}, M., {Rix},
  H.-W., {Ho}, L.~C., \& {Shields}, J.~C. 2002, \aj, 123, 1389

\bibitem[{{B{\"o}ker} {et~al.}(2004){B{\"o}ker}, {Sarzi}, {McLaughlin}, {van
  der Marel}, {Rix}, {Ho}, \& {Shields}}]{boker04}
{B{\"o}ker}, T., {Sarzi}, M., {McLaughlin}, D.~E., {van der Marel}, R.~P.,
  {Rix}, H.-W., {Ho}, L.~C., \& {Shields}, J.~C. 2004, \aj, 127, 105

\bibitem[{{B{\"o}ker} {et~al.}(1999){B{\"o}ker}, {van der Marel}, \&
  {Vacca}}]{boker99}
{B{\"o}ker}, T., {van der Marel}, R.~P., \& {Vacca}, W.~D. 1999, \aj, 118, 831

\bibitem[{{Bower} {et~al.}(1998){Bower}, {Green}, {Danks}, {Gull}, {Heap},
  {Hutchings}, {Joseph}, {Kaiser}, {Kimble}, {Kraemer}, {Weistrop}, {Woodgate},
  {Lindler}, {Hill}, {Malumuth}, {Baum}, {Sarajedini}, {Heckman}, {Wilson}, \&
  {Richstone}}]{bower98}
{Bower}, G.~A., {Green}, R.~F., {Danks}, A., {Gull}, T., {Heap}, S.,
  {Hutchings}, J., {Joseph}, C., {Kaiser}, M.~E., {Kimble}, R., {Kraemer}, S.,
  {Weistrop}, D., {Woodgate}, B., {Lindler}, D., {Hill}, R.~S., {Malumuth},
  E.~M., {Baum}, S., {Sarajedini}, V., {Heckman}, T.~M., {Wilson}, A.~S., \&
  {Richstone}, D.~O. 1998, \apjl, 492, L111+

\bibitem[{{Bower} {et~al.}(2000){Bower}, {Green}, {Quillen}, {Danks}, {Gull},
  {Hutchings}, {Joseph}, {Kaiser}, {Weistrop}, {Woodgate}, {Malumuth}, \&
  {Nelson}}]{bower00}
{Bower}, G.~A., {Green}, R.~F., {Quillen}, A.~C., {Danks}, A., {Gull}, T.,
  {Hutchings}, J., {Joseph}, C., {Kaiser}, M.~E., {Weistrop}, D., {Woodgate},
  B., {Malumuth}, E.~M., \& {Nelson}, C. 2000, \apj, 534, 189

\bibitem[{{Bromm} \& {Yoshida}(2011)}]{bromm11}
{Bromm}, V. \& {Yoshida}, N. 2011, \araa, 49, 373

\bibitem[{{Cappellari}(2002)}]{cappellari02}
{Cappellari}, M. 2002, \mnras, 333, 400

\bibitem[{{Cappellari}(2008)}]{cappellari08}
---. 2008, \mnras, 390, 71

\bibitem[{{Capuzzo-Dolcetta}(1993)}]{capuzzo93}
{Capuzzo-Dolcetta}, R. 1993, \apj, 415, 616

\bibitem[{{Carollo} {et~al.}(1998){Carollo}, {Stiavelli}, \&
  {Mack}}]{carollo98}
{Carollo}, C.~M., {Stiavelli}, M., \& {Mack}, J. 1998, \aj, 116, 68

\bibitem[{{C{\^o}t{\'e}} {et~al.}(2006){C{\^o}t{\'e}}, {Piatek}, {Ferrarese},
  {Jord{\'a}n}, {Merritt}, {Peng}, {Ha{\c s}egan}, {Blakeslee}, {Mei}, {West},
  {Milosavljevi{\'c}}, \& {Tonry}}]{cote06}
{C{\^o}t{\'e}}, P., {Piatek}, S., {Ferrarese}, L., {Jord{\'a}n}, A., {Merritt},
  D., {Peng}, E.~W., {Ha{\c s}egan}, M., {Blakeslee}, J.~P., {Mei}, S., {West},
  M.~J., {Milosavljevi{\'c}}, M., \& {Tonry}, J.~L. 2006, \apjs, 165, 57

\bibitem[{{Cretton} \& {van den Bosch}(1999)}]{cretton99}
{Cretton}, N. \& {van den Bosch}, F.~C. 1999, \apj, 514, 704

\bibitem[{{de Francesco} {et~al.}(2006){de Francesco}, {Capetti}, \&
  {Marconi}}]{defrancesco06}
{de Francesco}, G., {Capetti}, A., \& {Marconi}, A. 2006, \aap, 460, 439

\bibitem[{{Devereux} {et~al.}(2003){Devereux}, {Ford}, {Tsvetanov}, \&
  {Jacoby}}]{devereux03}
{Devereux}, N., {Ford}, H., {Tsvetanov}, Z., \& {Jacoby}, G. 2003, \aj, 125,
  1226

\bibitem[{{Di Matteo} {et~al.}(2005){Di Matteo}, {Springel}, \&
  {Hernquist}}]{dimatteo05}
{Di Matteo}, T., {Springel}, V., \& {Hernquist}, L. 2005, \nat, 433, 604

\bibitem[{{Ebisuzaki} {et~al.}(2001){Ebisuzaki}, {Makino}, {Tsuru}, {Funato},
  {Portegies Zwart}, {Hut}, {McMillan}, {Matsushita}, {Matsumoto}, \&
  {Kawabe}}]{ebisuzaki01}
{Ebisuzaki}, T., {Makino}, J., {Tsuru}, T.~G., {Funato}, Y., {Portegies Zwart},
  S., {Hut}, P., {McMillan}, S., {Matsushita}, S., {Matsumoto}, H., \&
  {Kawabe}, R. 2001, \apjl, 562, L19

\bibitem[{{Elmegreen} {et~al.}(2008){Elmegreen}, {Bournaud}, \&
  {Elmegreen}}]{elmegreen08}
{Elmegreen}, B.~G., {Bournaud}, F., \& {Elmegreen}, D.~M. 2008, \apj, 684, 829

\bibitem[{{Emsellem} {et~al.}(1994){Emsellem}, {Monnet}, \&
  {Bacon}}]{emsellem94}
{Emsellem}, E., {Monnet}, G., \& {Bacon}, R. 1994, \aap, 285, 723

\bibitem[{{Erwin} \& {Gadotti}(2010)}]{erwin10}
{Erwin}, P. \& {Gadotti}, D. 2010, in American Institute of Physics Conference
  Series, Vol. 1240, American Institute of Physics Conference Series, ed.
  {V.~P.~Debattista \& C.~C.~Popescu}, 223--226

\bibitem[{{Ferrarese} {et~al.}(2006a){Ferrarese}, {C{\^o}t{\'e}}, {Dalla
  Bont{\`a}}, {Peng}, {Merritt}, {Jord{\'a}n}, {Blakeslee}, {Ha{\c s}egan},
  {Mei}, {Piatek}, {Tonry}, \& {West}}]{ferrarese06a}
{Ferrarese}, L., {C{\^o}t{\'e}}, P., {Dalla Bont{\`a}}, E., {Peng}, E.~W.,
  {Merritt}, D., {Jord{\'a}n}, A., {Blakeslee}, J.~P., {Ha{\c s}egan}, M.,
  {Mei}, S., {Piatek}, S., {Tonry}, J.~L., \& {West}, M.~J. 2006a, \apjl, 644,
  L21

\bibitem[{{Ferrarese} {et~al.}(2006b){Ferrarese}, {C{\^o}t{\'e}}, {Jord{\'a}n},
  {Peng}, {Blakeslee}, {Piatek}, {Mei}, {Merritt}, {Milosavljevi{\'c}},
  {Tonry}, \& {West}}]{ferrarese06b}
{Ferrarese}, L., {C{\^o}t{\'e}}, P., {Jord{\'a}n}, A., {Peng}, E.~W.,
  {Blakeslee}, J.~P., {Piatek}, S., {Mei}, S., {Merritt}, D.,
  {Milosavljevi{\'c}}, M., {Tonry}, J.~L., \& {West}, M.~J. 2006b, \apjs, 164,
  334

\bibitem[{{Ferrarese} {et~al.}(1996){Ferrarese}, {Ford}, \&
  {Jaffe}}]{ferrarese96}
{Ferrarese}, L., {Ford}, H.~C., \& {Jaffe}, W. 1996, \apj, 470, 444

\bibitem[{{Ferrarese} \& {Merritt}(2000)}]{ferrarese00}
{Ferrarese}, L. \& {Merritt}, D. 2000, \apjl, 539, L9

\bibitem[{{Filippenko} \& {Ho}(2003)}]{filippenko03}
{Filippenko}, A.~V. \& {Ho}, L.~C. 2003, \apjl, 588, L13

\bibitem[{{Filippenko} \& {Sargent}(1989)}]{filippenko89}
{Filippenko}, A.~V. \& {Sargent}, W.~L.~W. 1989, \apjl, 342, L11

\bibitem[{{Freeman}(1993)}]{freeman93}
{Freeman}, K.~C. 1993, in Astronomical Society of the Pacific Conference
  Series, Vol.~48, The Globular Cluster-Galaxy Connection, ed. {G.~H.~Smith \&
  J.~P.~Brodie}, 608--+

\bibitem[{{Freitag} {et~al.}(2006{\natexlab{a}}){Freitag}, {G{\"u}rkan}, \&
  {Rasio}}]{freitag06b}
{Freitag}, M., {G{\"u}rkan}, M.~A., \& {Rasio}, F.~A. 2006{\natexlab{a}},
  \mnras, 368, 141

\bibitem[{{Freitag} {et~al.}(2006{\natexlab{b}}){Freitag}, {Rasio}, \&
  {Baumgardt}}]{freitag06a}
{Freitag}, M., {Rasio}, F.~A., \& {Baumgardt}, H. 2006{\natexlab{b}}, \mnras,
  368, 121

\bibitem[{{Gaburov} {et~al.}(2008){Gaburov}, {Gualandris}, \& {Portegies
  Zwart}}]{gaburov08}
{Gaburov}, E., {Gualandris}, A., \& {Portegies Zwart}, S. 2008, \mnras, 384,
  376

\bibitem[{{Ganda} {et~al.}(2009){Ganda}, {Peletier}, {Balcells}, \&
  {Falc{\'o}n-Barroso}}]{ganda09}
{Ganda}, K., {Peletier}, R.~F., {Balcells}, M., \& {Falc{\'o}n-Barroso}, J.
  2009, \mnras, 395, 1669

\bibitem[{{Gebhardt} {et~al.}(2011){Gebhardt}, {Adams}, {Richstone}, {Lauer},
  {Faber}, {G{\"u}ltekin}, {Murphy}, \& {Tremaine}}]{gebhardt11}
{Gebhardt}, K., {Adams}, J., {Richstone}, D., {Lauer}, T.~R., {Faber}, S.~M.,
  {G{\"u}ltekin}, K., {Murphy}, J., \& {Tremaine}, S. 2011, \apj, 729, 119

\bibitem[{{Gebhardt} {et~al.}(2000a){Gebhardt}, {Bender}, {Bower}, {Dressler},
  {Faber}, {Filippenko}, {Green}, {Grillmair}, {Ho}, {Kormendy}, {Lauer},
  {Magorrian}, {Pinkney}, {Richstone}, \& {Tremaine}}]{gebhardt00a}
{Gebhardt}, K., {Bender}, R., {Bower}, G., {Dressler}, A., {Faber}, S.~M.,
  {Filippenko}, A.~V., {Green}, R., {Grillmair}, C., {Ho}, L.~C., {Kormendy},
  J., {Lauer}, T.~R., {Magorrian}, J., {Pinkney}, J., {Richstone}, D., \&
  {Tremaine}, S. 2000a, \apjl, 539, L13

\bibitem[{{Gebhardt} {et~al.}(2001){Gebhardt}, {Lauer}, {Kormendy}, {Pinkney},
  {Bower}, {Green}, {Gull}, {Hutchings}, {Kaiser}, {Nelson}, {Richstone}, \&
  {Weistrop}}]{gebhardt01}
{Gebhardt}, K., {Lauer}, T.~R., {Kormendy}, J., {Pinkney}, J., {Bower}, G.~A.,
  {Green}, R., {Gull}, T., {Hutchings}, J.~B., {Kaiser}, M.~E., {Nelson},
  C.~H., {Richstone}, D., \& {Weistrop}, D. 2001, \aj, 122, 2469

\bibitem[{{Gebhardt} {et~al.}(2005){Gebhardt}, {Rich}, \& {Ho}}]{gebhardt05}
{Gebhardt}, K., {Rich}, R.~M., \& {Ho}, L.~C. 2005, \apj, 634, 1093

\bibitem[{{Gebhardt} {et~al.}(2000b){Gebhardt}, {Richstone}, {Kormendy},
  {Lauer}, {Ajhar}, {Bender}, {Dressler}, {Faber}, {Grillmair}, {Magorrian}, \&
  {Tremaine}}]{gebhardt00b}
{Gebhardt}, K., {Richstone}, D., {Kormendy}, J., {Lauer}, T.~R., {Ajhar},
  E.~A., {Bender}, R., {Dressler}, A., {Faber}, S.~M., {Grillmair}, C.,
  {Magorrian}, J., \& {Tremaine}, S. 2000b, \aj, 119, 1157

\bibitem[{{Gebhardt} {et~al.}(2003){Gebhardt}, {Richstone}, {Tremaine},
  {Lauer}, {Bender}, {Bower}, {Dressler}, {Faber}, {Filippenko}, {Green},
  {Grillmair}, {Ho}, {Kormendy}, {Magorrian}, \& {Pinkney}}]{gebhardt03}
{Gebhardt}, K., {Richstone}, D., {Tremaine}, S., {Lauer}, T.~R., {Bender}, R.,
  {Bower}, G., {Dressler}, A., {Faber}, S.~M., {Filippenko}, A.~V., {Green},
  R., {Grillmair}, C., {Ho}, L.~C., {Kormendy}, J., {Magorrian}, J., \&
  {Pinkney}, J. 2003, \apj, 583, 92

\bibitem[{{Gebhardt} \& {Thomas}(2009)}]{gebhardt09}
{Gebhardt}, K. \& {Thomas}, J. 2009, \apj, 700, 1690

\bibitem[{{Genzel} {et~al.}(2010){Genzel}, {Eisenhauer}, \&
  {Gillessen}}]{genzel10}
{Genzel}, R., {Eisenhauer}, F., \& {Gillessen}, S. 2010, Reviews of Modern
  Physics, 82, 3121

\bibitem[{{Glebbeek} {et~al.}(2009){Glebbeek}, {Gaburov}, {de Mink}, {Pols}, \&
  {Portegies Zwart}}]{glebbeek09}
{Glebbeek}, E., {Gaburov}, E., {de Mink}, S.~E., {Pols}, O.~R., \& {Portegies
  Zwart}, S.~F. 2009, \aap, 497, 255

\bibitem[{{Gliozzi} {et~al.}(2009){Gliozzi}, {Satyapal}, {Eracleous},
  {Titarchuk}, \& {Cheung}}]{gliozzi09}
{Gliozzi}, M., {Satyapal}, S., {Eracleous}, M., {Titarchuk}, L., \& {Cheung},
  C.~C. 2009, \apj, 700, 1759

\bibitem[{{Graham} \& {Driver}(2007)}]{graham07}
{Graham}, A.~W. \& {Driver}, S.~P. 2007, \apj, 655, 77

\bibitem[{{Graham} {et~al.}(2011){Graham}, {Onken}, {Athanassoula}, \&
  {Combes}}]{graham11}
{Graham}, A.~W., {Onken}, C.~A., {Athanassoula}, E., \& {Combes}, F. 2011,
  \mnras, 412, 2211

\bibitem[{{Graham} \& {Spitler}(2009)}]{graham09b}
{Graham}, A.~W. \& {Spitler}, L.~R. 2009, \mnras, 397, 2148

\bibitem[{{Greene} \& {Ho}(2007)}]{greene07}
{Greene}, J.~E. \& {Ho}, L.~C. 2007, \apj, 670, 92

\bibitem[{{Greene} {et~al.}(2008){Greene}, {Ho}, \& {Barth}}]{greene08}
{Greene}, J.~E., {Ho}, L.~C., \& {Barth}, A.~J. 2008, \apj, 688, 159

\bibitem[{{Greene} {et~al.}(2010){Greene}, {Peng}, {Kim}, {Kuo}, {Braatz},
  {Violette Impellizzeri}, {Condon}, {Lo}, {Henkel}, \& {Reid}}]{greene10}
{Greene}, J.~E., {Peng}, C.~Y., {Kim}, M., {Kuo}, C.-Y., {Braatz}, J.~A.,
  {Violette Impellizzeri}, C.~M., {Condon}, J.~J., {Lo}, K.~Y., {Henkel}, C.,
  \& {Reid}, M.~J. 2010, \apj, 721, 26

\bibitem[{{G{\"u}ltekin} {et~al.}(2009){G{\"u}ltekin}, {Richstone}, {Gebhardt},
  {Lauer}, {Tremaine}, {Aller}, {Bender}, {Dressler}, {Faber}, {Filippenko},
  {Green}, {Ho}, {Kormendy}, {Magorrian}, {Pinkney}, \& {Siopis}}]{gueltekin09}
{G{\"u}ltekin}, K., {Richstone}, D.~O., {Gebhardt}, K., {Lauer}, T.~R.,
  {Tremaine}, S., {Aller}, M.~C., {Bender}, R., {Dressler}, A., {Faber}, S.~M.,
  {Filippenko}, A.~V., {Green}, R., {Ho}, L.~C., {Kormendy}, J., {Magorrian},
  J., {Pinkney}, J., \& {Siopis}, C. 2009, \apj, 698, 198

\bibitem[{{G{\"u}ltekin} {et~al.}(2011{\natexlab{a}}){G{\"u}ltekin},
  {Tremaine}, {Loeb}, \& {Richstone}}]{gultekin11}
{G{\"u}ltekin}, K., {Tremaine}, S., {Loeb}, A., \& {Richstone}, D.~O.
  2011{\natexlab{a}}, \apj, 738, 17

\bibitem[{{G{\"u}ltekin} {et~al.}(2011{\natexlab{b}}){G{\"u}ltekin},
  {Tremaine}, {Loeb}, \& {Richstone}}]{jalali11}
---. 2011{\natexlab{b}}, \apj, 738, 17

\bibitem[{{G{\"u}rkan} {et~al.}(2004){G{\"u}rkan}, {Freitag}, \&
  {Rasio}}]{gurkan04}
{G{\"u}rkan}, M.~A., {Freitag}, M., \& {Rasio}, F.~A. 2004, \apj, 604, 632

\bibitem[{{H\"aring} \& {Rix}(2004)}]{haring04}
{H\"aring}, N. \& {Rix}, H.-W. 2004, \apjl, 604, L89

\bibitem[{{Hartmann} {et~al.}(2011){Hartmann}, {Debattista}, {Seth},
  {Cappellari}, \& {Quinn}}]{hartmann11}
{Hartmann}, M., {Debattista}, V.~P., {Seth}, A., {Cappellari}, M., \& {Quinn},
  T.~R. 2011, \mnras, 418, 2697

\bibitem[{{Hopkins} {et~al.}(2006){Hopkins}, {Hernquist}, {Cox}, {Di Matteo},
  {Robertson}, \& {Springel}}]{hopkins06}
{Hopkins}, P.~F., {Hernquist}, L., {Cox}, T.~J., {Di Matteo}, T., {Robertson},
  B., \& {Springel}, V. 2006, \apjs, 163, 1

\bibitem[{{Jahnke} \& {Macci{\`o}}(2011)}]{jahnke11}
{Jahnke}, K. \& {Macci{\`o}}, A.~V. 2011, \apj, 734, 92

\bibitem[{{Jee} {et~al.}(2007){Jee}, {Blakeslee}, {Sirianni}, {Martel},
  {White}, \& {Ford}}]{jee07}
{Jee}, M.~J., {Blakeslee}, J.~P., {Sirianni}, M., {Martel}, A.~R., {White},
  R.~L., \& {Ford}, H.~C. 2007, \pasp, 119, 1403

\bibitem[{{Koleva} {et~al.}(2011){Koleva}, {Prugniel}, {de Rijcke}, \&
  {Zeilinger}}]{koleva11}
{Koleva}, M., {Prugniel}, P., {de Rijcke}, S., \& {Zeilinger}, W.~W. 2011,
  \mnras, 417, 1643

\bibitem[{{Kormendy} {et~al.}(2011){Kormendy}, {Bender}, \&
  {Cornell}}]{kormendy11a}
{Kormendy}, J., {Bender}, R., \& {Cornell}, M.~E. 2011, \nat, 469, 374

\bibitem[{{Krist}(1995)}]{krist95}
{Krist}, J. 1995, in Astronomical Society of the Pacific Conference Series,
  Vol.~77, Astronomical Data Analysis Software and Systems IV, ed. {R.~A.~Shaw,
  H.~E.~Payne, \& J.~J.~E.~Hayes}, 349--+

\bibitem[{{Lauer} {et~al.}(2007){Lauer}, {Gebhardt}, {Faber}, {Richstone},
  {Tremaine}, {Kormendy}, {Aller}, {Bender}, {Dressler}, {Filippenko}, {Green},
  \& {Ho}}]{lauer07}
{Lauer}, T.~R., {Gebhardt}, K., {Faber}, S.~M., {Richstone}, D., {Tremaine},
  S., {Kormendy}, J., {Aller}, M.~C., {Bender}, R., {Dressler}, A.,
  {Filippenko}, A.~V., {Green}, R., \& {Ho}, L.~C. 2007, \apj, 664, 226

\bibitem[{{L{\"u}tzgendorf} {et~al.}(2011){L{\"u}tzgendorf}, {Kissler-Patig},
  {Noyola}, {Jalali}, {de Zeeuw}, {Gebhardt}, \& {Baumgardt}}]{luetzgendorf11}
{L{\"u}tzgendorf}, N., {Kissler-Patig}, M., {Noyola}, E., {Jalali}, B., {de
  Zeeuw}, P.~T., {Gebhardt}, K., \& {Baumgardt}, H. 2011, \aap, 533, A36

\bibitem[{{Mayer} {et~al.}(2010){Mayer}, {Kazantzidis}, {Escala}, \&
  {Callegari}}]{mayer10}
{Mayer}, L., {Kazantzidis}, S., {Escala}, A., \& {Callegari}, S. 2010, \nat,
  466, 1082

\bibitem[{{McAlpine} {et~al.}(2011){McAlpine}, {Satyapal}, {Gliozzi}, {Cheung},
  {Sambruna}, \& {Eracleous}}]{mcalpine11}
{McAlpine}, W., {Satyapal}, S., {Gliozzi}, M., {Cheung}, C.~C., {Sambruna},
  R.~M., \& {Eracleous}, M. 2011, \apj, 728, 25

\bibitem[{{McLaughlin}(1995)}]{mclaughlin95}
{McLaughlin}, D.~E. 1995, \aj, 109, 2034

\bibitem[{{Merritt}(2006b)}]{merritt06b}
{Merritt}, D. 2006b, Reports on Progress in Physics, 69, 2513

\bibitem[{{Merritt} {et~al.}(2001){Merritt}, {Ferrarese}, \&
  {Joseph}}]{merritt01}
{Merritt}, D., {Ferrarese}, L., \& {Joseph}, C.~L. 2001, Science, 293, 1116

\bibitem[{{Merritt} \& {Szell}(2006a)}]{merritt06a}
{Merritt}, D. \& {Szell}, A. 2006a, \apj, 648, 890

\bibitem[{{Milosavljevi{\'c}}(2004)}]{milosavljevic04}
{Milosavljevi{\'c}}, M. 2004, \apjl, 605, L13

\bibitem[{{Miocchi} {et~al.}(2006){Miocchi}, {Capuzzo Dolcetta}, {Di Matteo},
  \& {Vicari}}]{miocchi06}
{Miocchi}, P., {Capuzzo Dolcetta}, R., {Di Matteo}, P., \& {Vicari}, A. 2006,
  \apj, 644, 940

\bibitem[{{Nayakshin} {et~al.}(2009){Nayakshin}, {Wilkinson}, \&
  {King}}]{nayakshin09}
{Nayakshin}, S., {Wilkinson}, M.~I., \& {King}, A. 2009, \mnras, 398, L54

\bibitem[{{Neumayer} {et~al.}(2011){Neumayer}, {Walcher}, {Andersen},
  {S{\'a}nchez}, {B{\"o}ker}, \& {Rix}}]{neumayer11}
{Neumayer}, N., {Walcher}, C.~J., {Andersen}, D., {S{\'a}nchez}, S.~F.,
  {B{\"o}ker}, T., \& {Rix}, H.-W. 2011, \mnras, 413, 1875

\bibitem[{{Noyola} {et~al.}(2008){Noyola}, {Gebhardt}, \&
  {Bergmann}}]{noyola08}
{Noyola}, E., {Gebhardt}, K., \& {Bergmann}, M. 2008, \apj, 676, 1008

\bibitem[{{Noyola} {et~al.}(2010){Noyola}, {Gebhardt}, {Kissler-Patig},
  {L{\"u}tzgendorf}, {Jalali}, {de Zeeuw}, \& {Baumgardt}}]{noyola10}
{Noyola}, E., {Gebhardt}, K., {Kissler-Patig}, M., {L{\"u}tzgendorf}, N.,
  {Jalali}, B., {de Zeeuw}, P.~T., \& {Baumgardt}, H. 2010, \apjl, 719, L60

\bibitem[{{Peng}(2007)}]{peng07}
{Peng}, C.~Y. 2007, \apj, 671, 1098

\bibitem[{{Peng} {et~al.}(2002){Peng}, {Ho}, {Impey}, \& {Rix}}]{peng02}
{Peng}, C.~Y., {Ho}, L.~C., {Impey}, C.~D., \& {Rix}, H.-W. 2002, \aj, 124, 266

\bibitem[{{Peterson} {et~al.}(2005){Peterson}, {Bentz}, {Desroches},
  {Filippenko}, {Ho}, {Kaspi}, {Laor}, {Maoz}, {Moran}, {Pogge}, \&
  {Quillen}}]{peterson05}
{Peterson}, B.~M., {Bentz}, M.~C., {Desroches}, L.-B., {Filippenko}, A.~V.,
  {Ho}, L.~C., {Kaspi}, S., {Laor}, A., {Maoz}, D., {Moran}, E.~C., {Pogge},
  R.~W., \& {Quillen}, A.~C. 2005, \apj, 632, 799

\bibitem[{{Pflamm-Altenburg} \& {Kroupa}(2009)}]{pflamm09}
{Pflamm-Altenburg}, J. \& {Kroupa}, P. 2009, \mnras, 397, 488

\bibitem[{{Portegies Zwart} {et~al.}(2004){Portegies Zwart}, {Baumgardt},
  {Hut}, {Makino}, \& {McMillan}}]{portegies04}
{Portegies Zwart}, S.~F., {Baumgardt}, H., {Hut}, P., {Makino}, J., \&
  {McMillan}, S.~L.~W. 2004, \nat, 428, 724

\bibitem[{{Regan} \& {Haehnelt}(2009)}]{regan09}
{Regan}, J.~A. \& {Haehnelt}, M.~G. 2009, \mnras, 396, 343

\bibitem[{{Rossa} {et~al.}(2006){Rossa}, {van der Marel}, {B{\"o}ker},
  {Gerssen}, {Ho}, {Rix}, {Shields}, \& {Walcher}}]{rossa06}
{Rossa}, J., {van der Marel}, R.~P., {B{\"o}ker}, T., {Gerssen}, J., {Ho},
  L.~C., {Rix}, H.-W., {Shields}, J.~C., \& {Walcher}, C.-J. 2006, \aj, 132,
  1074

\bibitem[{{Rusli} {et~al.}(2011){Rusli}, {Thomas}, {Erwin}, {Saglia}, {Nowak},
  \& {Bender}}]{rusli11}
{Rusli}, S.~P., {Thomas}, J., {Erwin}, P., {Saglia}, R.~P., {Nowak}, N., \&
  {Bender}, R. 2011, \mnras, 410, 1223

\bibitem[{{Sarzi} {et~al.}(2001){Sarzi}, {Rix}, {Shields}, {Rudnick}, {Ho},
  {McIntosh}, {Filippenko}, \& {Sargent}}]{sarzi01}
{Sarzi}, M., {Rix}, H.-W., {Shields}, J.~C., {Rudnick}, G., {Ho}, L.~C.,
  {McIntosh}, D.~H., {Filippenko}, A.~V., \& {Sargent}, W.~L.~W. 2001, \apj,
  550, 65

\bibitem[{{Satyapal} {et~al.}(2008){Satyapal}, {Vega}, {Dudik}, {Abel}, \&
  {Heckman}}]{satyapal08}
{Satyapal}, S., {Vega}, D., {Dudik}, R.~P., {Abel}, N.~P., \& {Heckman}, T.
  2008, \apj, 677, 926

\bibitem[{{Sch{\"o}del} {et~al.}(2007){Sch{\"o}del}, {Eckart}, {Alexander},
  {Merritt}, {Genzel}, {Sternberg}, {Meyer}, {Kul}, {Moultaka}, {Ott}, \&
  {Straubmeier}}]{schoedel07}
{Sch{\"o}del}, R., {Eckart}, A., {Alexander}, T., {Merritt}, D., {Genzel}, R.,
  {Sternberg}, A., {Meyer}, L., {Kul}, F., {Moultaka}, J., {Ott}, T., \&
  {Straubmeier}, C. 2007, \aap, 469, 125

\bibitem[{{Scorza} \& {van den Bosch}(1998)}]{scorza98}
{Scorza}, C. \& {van den Bosch}, F.~C. 1998, \mnras, 300, 469

\bibitem[{{Serra} \& {Trager}(2007)}]{serra07}
{Serra}, P. \& {Trager}, S.~C. 2007, \mnras, 374, 769

\bibitem[{{Seth} {et~al.}(2008){Seth}, {Ag{\"u}eros}, {Lee}, \&
  {Basu-Zych}}]{seth08}
{Seth}, A., {Ag{\"u}eros}, M., {Lee}, D., \& {Basu-Zych}, A. 2008, \apj, 678,
  116

\bibitem[{{Seth} {et~al.}(2010){Seth}, {Cappellari}, {Neumayer}, {Caldwell},
  {Bastian}, {Olsen}, {Blum}, {Debattista}, {McDermid}, {Puzia}, \&
  {Stephens}}]{seth10}
{Seth}, A.~C., {Cappellari}, M., {Neumayer}, N., {Caldwell}, N., {Bastian}, N.,
  {Olsen}, K., {Blum}, R.~D., {Debattista}, V.~P., {McDermid}, R., {Puzia}, T.,
  \& {Stephens}, A. 2010, \apj, 714, 713

\bibitem[{{Seth} {et~al.}(2006){Seth}, {Dalcanton}, {Hodge}, \&
  {Debattista}}]{seth06}
{Seth}, A.~C., {Dalcanton}, J.~J., {Hodge}, P.~W., \& {Debattista}, V.~P. 2006,
  \aj, 132, 2539

\bibitem[{{Shapiro} {et~al.}(2006){Shapiro}, {Cappellari}, {de Zeeuw},
  {McDermid}, {Gebhardt}, {van den Bosch}, \& {Statler}}]{shapiro06}
{Shapiro}, K.~L., {Cappellari}, M., {de Zeeuw}, T., {McDermid}, R.~M.,
  {Gebhardt}, K., {van den Bosch}, R.~C.~E., \& {Statler}, T.~S. 2006, \mnras,
  370, 559

\bibitem[{{Shields} {et~al.}(2008){Shields}, {Walcher}, {B{\"o}ker}, {Ho},
  {Rix}, \& {van der Marel}}]{shields08}
{Shields}, J.~C., {Walcher}, C.~J., {B{\"o}ker}, T., {Ho}, L.~C., {Rix}, H.-W.,
  \& {van der Marel}, R.~P. 2008, \apj, 682, 104

\bibitem[{{Tremaine} {et~al.}(1975){Tremaine}, {Ostriker}, \&
  {Spitzer}}]{tremaine75}
{Tremaine}, S.~D., {Ostriker}, J.~P., \& {Spitzer}, Jr., L. 1975, \apj, 196,
  407

\bibitem[{{Valluri} {et~al.}(2005){Valluri}, {Ferrarese}, {Merritt}, \&
  {Joseph}}]{valluri05}
{Valluri}, M., {Ferrarese}, L., {Merritt}, D., \& {Joseph}, C.~L. 2005, \apj,
  628, 137

\bibitem[{{van der Marel} \& {Anderson}(2010)}]{marel10}
{van der Marel}, R.~P. \& {Anderson}, J. 2010, \apj, 710, 1063

\bibitem[{{Volonteri} {et~al.}(2008){Volonteri}, {Lodato}, \&
  {Natarajan}}]{volonteri08}
{Volonteri}, M., {Lodato}, G., \& {Natarajan}, P. 2008, \mnras, 383, 1079

\bibitem[{{Walcher} {et~al.}(2006){Walcher}, {B{\"o}ker}, {Charlot}, {Ho},
  {Rix}, {Rossa}, {Shields}, \& {van der Marel}}]{walcher06}
{Walcher}, C.~J., {B{\"o}ker}, T., {Charlot}, S., {Ho}, L.~C., {Rix}, H.-W.,
  {Rossa}, J., {Shields}, J.~C., \& {van der Marel}, R.~P. 2006, \apj, 649, 692

\bibitem[{{Walcher} {et~al.}(2005){Walcher}, {van der Marel}, {McLaughlin},
  {Rix}, {B{\"o}ker}, {H{\"a}ring}, {Ho}, {Sarzi}, \& {Shields}}]{walcher05}
{Walcher}, C.~J., {van der Marel}, R.~P., {McLaughlin}, D., {Rix}, H.-W.,
  {B{\"o}ker}, T., {H{\"a}ring}, N., {Ho}, L.~C., {Sarzi}, M., \& {Shields},
  J.~C. 2005, \apj, 618, 237

\bibitem[{{Walsh} {et~al.}(2010){Walsh}, {Barth}, \& {Sarzi}}]{walsh10}
{Walsh}, J.~L., {Barth}, A.~J., \& {Sarzi}, M. 2010, \apj, 721, 762

\bibitem[{{Wehner} \& {Harris}(2006)}]{wehner06}
{Wehner}, E.~H. \& {Harris}, W.~E. 2006, \apjl, 644, L17

\bibitem[{{Weinzirl} {et~al.}(2009){Weinzirl}, {Jogee}, {Khochfar}, {Burkert},
  \& {Kormendy}}]{weinzirl09}
{Weinzirl}, T., {Jogee}, S., {Khochfar}, S., {Burkert}, A., \& {Kormendy}, J.
  2009, \apj, 696, 411

\bibitem[{{Young} {et~al.}(1978){Young}, {Westphal}, {Kristian}, {Wilson}, \&
  {Landauer}}]{young78}
{Young}, P.~J., {Westphal}, J.~A., {Kristian}, J., {Wilson}, C.~P., \&
  {Landauer}, F.~P. 1978, \apj, 221, 721

\bibitem[{{Zinnecker} {et~al.}(1988){Zinnecker}, {Keable}, {Dunlop}, {Cannon},
  \& {Griffiths}}]{zinnecker88}
{Zinnecker}, H., {Keable}, C.~J., {Dunlop}, J.~S., {Cannon}, R.~D., \&
  {Griffiths}, W.~K. 1988, in IAU Symposium, Vol. 126, The Harlow-Shapley
  Symposium on Globular Cluster Systems in Galaxies, ed. {J.~E.~Grindlay \&
  A.~G.~D.~Philip}, 603--+

\end{thebibliography}

\end{document}